\documentclass[10pt,journal,compsoc]{IEEEtran}
\usepackage{hyperref}
\usepackage{url}
\usepackage{color}
\usepackage{subfigure}
\usepackage{graphicx}
\usepackage{multirow}
\usepackage{array}
\usepackage{booktabs}
\usepackage{ulem}
\usepackage{enumitem}
\usepackage{algpseudocode}
\usepackage{algorithmicx,algorithm}

\long\def\comment#1{}
\usepackage{amsmath,amsfonts,bm}
\usepackage{bbm}

\newcommand{\ie}{\textit{i.e.}}
\newcommand{\eg}{\textit{e.g.}}

\ifCLASSOPTIONcompsoc
  \usepackage[nocompress]{cite}
\else
  \usepackage{cite}
\fi
\ifCLASSINFOpdf
\else
\fi
\begin{document}
%
\title{Versatile Weight Attack via Flipping Limited Bits}
%
%
%
%

\author{Jiawang~Bai,
        Baoyuan~Wu,
        Zhifeng~Li,
        ~and~Shu-tao~Xia
\IEEEcompsocitemizethanks{
\IEEEcompsocthanksitem Jiawang Bai and Shu-tao Xia are with the Tsinghua Shenzhen International Graduate School, Tsinghua University, Shenzhen 518057, China. \protect\\
E-mail: \{bjw19@mails.tsinghua.edu.cn, xiast@sz.tsinghua.edu.cn\}
\IEEEcompsocthanksitem Baoyuan Wu is with the School of Data Science, Chinese University of Hong Kong, Shenzhen, and the Secure Computing Lab of Big Data, Shenzhen Research Institute of Big Data, 518100, China. \protect\\
E-mail: wubaoyuan@cuhk.edu.cn
\IEEEcompsocthanksitem Zhifeng Li is with Tencent Data Platform, Shenzhen 518057, China. \protect\\
E-mail: michaelzfli@tencent.com
\IEEEcompsocthanksitem Corresponding authors: Baoyuan Wu and Shu-tao Xia}

}

\IEEEtitleabstractindextext{%
\begin{abstract}
To explore the vulnerability of deep neural networks (DNNs), many attack paradigms have been well studied, such as the poisoning-based backdoor attack in the training stage and the adversarial attack in the inference stage. In this paper, we study a novel attack paradigm, which modifies model parameters in the deployment stage. Considering the effectiveness and stealthiness goals, we provide a general formulation to perform the bit-flip based weight attack, where the effectiveness term could be customized depending on the attacker's purpose. Furthermore, we present two cases of the general formulation with different malicious purposes, $i.e.$, single sample attack (SSA) and triggered samples attack (TSA). To this end, we formulate this problem as a mixed integer programming (MIP) to jointly determine the state of the binary bits (0 or 1) in the memory and learn the sample modification. Utilizing the latest technique in integer programming, we equivalently reformulate this MIP problem as a continuous optimization problem, which can be effectively and efficiently solved using the alternating direction method of multipliers (ADMM) method. Consequently, the flipped critical bits can be easily determined through optimization, rather than using a heuristic strategy. Extensive experiments demonstrate the superiority of SSA and TSA in attacking DNNs.
\end{abstract}

}

\maketitle

\IEEEdisplaynontitleabstractindextext

%
\IEEEpeerreviewmaketitle





\IEEEraisesectionheading{\section{Introduction}\label{sec:introduction}}

\IEEEPARstart{D}EEP neural networks (DNNs) have achieved state-of-the-art performance in many applications, including computer vision \cite{he2016deep}, natural language processing \cite{zhang2018neural}, and robotic manipulation \cite{levine2016end}.
However, many works \cite{szegedy2013intriguing,gu2019badnets} have revealed that DNNs are vulnerable to a range of attacks, which has attracted great attention, especially for security-critical applications (\eg, face recognition \cite{dong2019efficient}, medical diagnosis \cite{finlayson2019adversarial}, and autonomous driving \cite{eykholt2018robust,arnab2019robustness}). 
For example, backdoor attack \cite{SahaSP20,xie2019dba} manipulates the behavior of the DNN model by mainly poisoning some training data in the training stage; adversarial attack \cite{goodfellow2014explaining,moosavi2017universal} aims to fool the DNN model by adding malicious imperceptible perturbations onto the input in the inference stage.

\begin{figure}[t]
    \centering
    \includegraphics[width=0.47\textwidth]{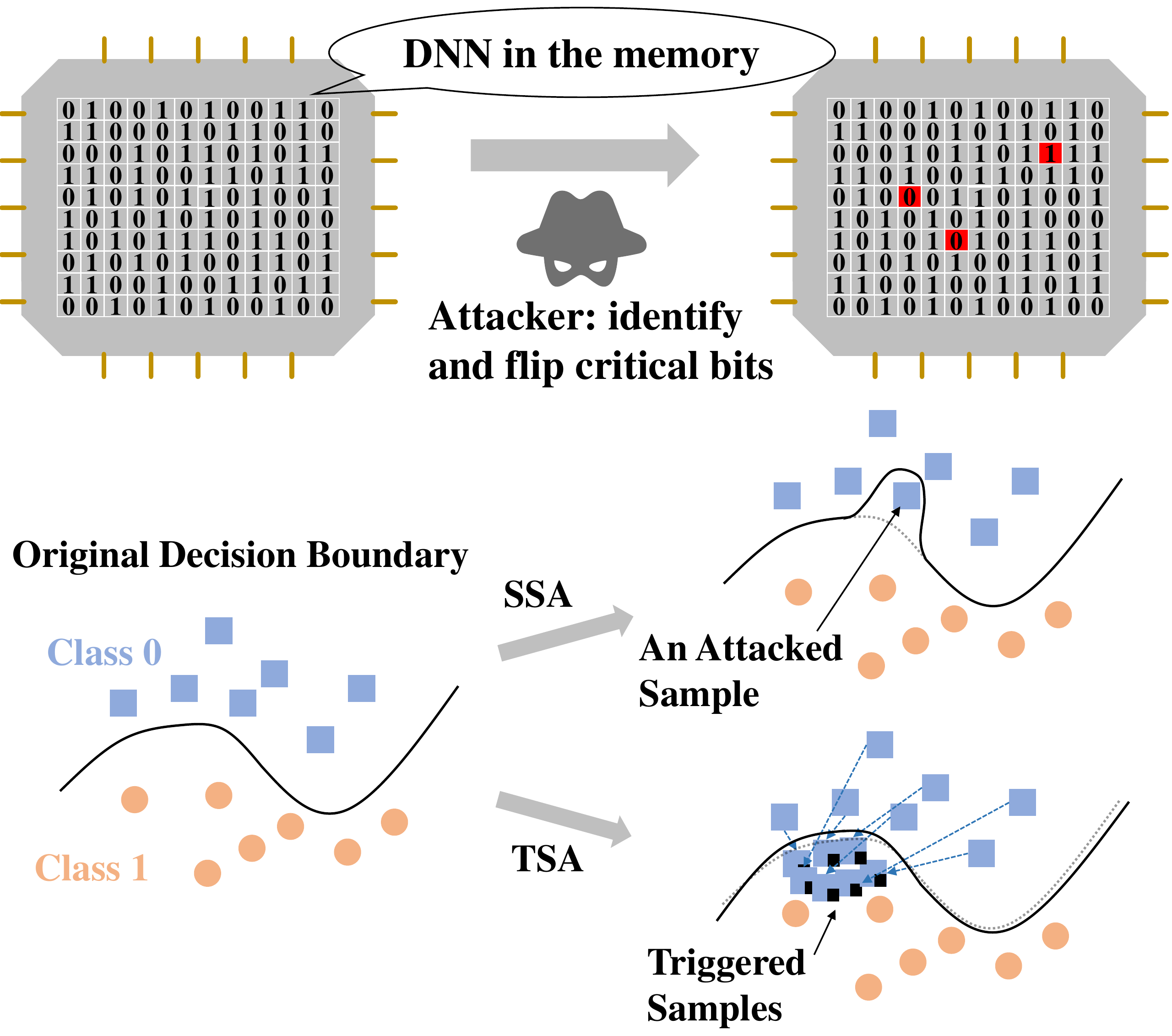}
    \caption{Demonstration of our proposed attack against a deployed DNN in the memory. By flipping critical bits (marked in red), our method can achieve some malicious purposes. For example, single sample attack (SSA) can manipulate the behavior of the DNN on an attacked sample without any sample modification while making little difference to other samples; triggered samples attack (TSA) can mislead the triggered samples into the target class while not significantly reducing the prediction accuracy of original samples.}
	\label{fig:demo}
\end{figure}

Besides the above backdoor attack and adversarial attack, \textit{weight attack} \cite{breier2018practical}, as a novel attack paradigm, has not been well studied. It assumes that the attacker has full access to the memory of a device, such that he/she can directly change the parameters of a deployed model to achieve some malicious purposes. For example, Rakin $et$ $al.$ \cite{rakin2019bit} proposed to crush a fully functional DNN model by converting it to a random output generator. 
Since the weight attack neither modifies the training data nor controls the training process and performs in the deployment stage, it is difficult for both the service provider and the user to realize the existence of the attack. 
In practice, since the deployed DNN model is stored as binary bits in the memory, the attacker can modify the model parameters using some physical fault injection techniques, such as Row Hammer Attack \cite{agoyan2010flip,selmke2015precise} and Laser Beam Attack \cite{kim2014flipping}. These techniques can precisely flip any bit of the data in the memory.
Some previous works \cite{rakin2019bit,rakin2020tbt,rakin2020t} have demonstrated that it is feasible to change the model weights via flipping weight bits to achieve some malicious purposes. Their efforts partially focused on identifying the critical bits in a large number of model parameters.
However, 
these methods mostly used some heuristic strategies. For example, the attack in \cite{rakin2019bit} combined gradient ranking and progressive search to identify the critical bits for flipping.


This work also focuses on the bit-level weight attack against DNNs in the deployment stage, whereas with two different goals, including {\it effectiveness} and {\it stealthiness}. The effectiveness requires that the attacked DNN can meet the attacker-specified malicious purpose, while the stealthiness encourages that the attacked DNN behaves normally on samples except the attacked sample(s). We formulate these two goals as two terms in the objective function, respectively, where the effectiveness loss can be customized depending on the attacker's purpose. This general formulation results in a versatile weight attack.
Specifically, we treat each bit in the memory as a binary variable and the modification on the samples as a continuous operator, and our task is to determine the state of each bit (\ie, 0 or 1) and the learnable parameter of the sample modification. Accordingly, it can be formulated as a mixed integer programming (MIP) problem. 
To further improve the stealthiness, we also limit the number of flipped bits, which can be formulated as a cardinality constraint. 
%

Considering practical attack scenarios, we specify the general formulation as two types of attack:  single sample attack (SSA) and triggered samples attack (TSA), by introducing different effectiveness goals. 
SSA aims at misclassifying a specific sample to a predefined target class without any sample modification, while TSA is to misclassify the samples embedded with a designed trigger. Besides, both attacks expect that the attacked DNN can meet the stealthiness goal to make the attack undetectable. Their goals are demonstrated in Fig. \ref{fig:demo}. They are special cases of the above general formulation and can be formulated as MIP problems.

However, how to solve the MIP problem with a cardinality constraint is a challenging problem. 
Fortunately, inspired by an advanced optimization method, the $\ell_p$-box ADMM \cite{wu2018ell}, this problem can be reformulated as a continuous optimization problem, which can further be efficiently and effectively solved by the alternating direction method of multipliers (ADMM) \cite{glowinski1975approximation,gabay1976dual}. 
Consequently, the flipped bits can be determined through optimization rather than the heuristic strategy, which makes our attack more effective.
Note that we also conduct experiments against the quantized DNN models, following the setting in some related works \cite{rakin2019bit,rakin2020tbt}.
Extensive experiments demonstrate the superiority of the proposed method over several existing weight attacks \footnote{Our code is available at: \url{https://github.com/jiawangbai/Versatile-Weight-Attack}}.
For example, in attacking an 8-bit quantized ResNet-18 model on ImageNet, on average,  SSA achieves a 100\% attack success rate with 7.37 bit-flips and 0.09\% accuracy degradation of the rest unspecific inputs, and TSA achieves 95.63\% attack success rate with 3.4 bit-flips and 0.05\% accuracy degradation of the original samples. Moreover, we also demonstrate that the proposed method is also more resistant to existing defense methods. 

This work builds upon the preliminary conference paper \cite{bai2021targeted}, which primarily focuses on attacking a specified sample without modifying the attacked sample. 
The new major contributions are summarized as follows. 
\begin{itemize}
\item[1)] We expand the attack in the conference paper to a general formulation, which can be flexibly specified to apply in different attack scenarios. This general formulation consists of identifying the flipped bits and learning the modification on the samples, resulting in a MIP problem.
\item[2)] Based on the general formulation, we propose a new type of attack, $i.e.$, TSA. Compared to SSA, TSA aims at misclassifying all samples with the learned trigger, which may be suitable to the scenarios that require various malicious inputs. 
\item[3)] Utilizing the general framework for solving this MIP problem, we conduct more experiments and ablation studies to verify the proposed attacks. In particular, in the new attack scenario, TSA reduces the number of bit-flips greatly compared to state-of-the-art methods.
\end{itemize}

\section{Related Work}
\label{sec:related_work}

Our work is related to various attack paradigms that explore the vulnerability of DNNs, including adversarial attack, backdoor attack, and weight attack. We compare these three attack paradigms from three aspects in Table \ref{tab:attack_compare} and discuss them briefly in what follows.

\begin{table}[t]
\centering
\caption{Comparison between three attack paradigms in terms of attack stage, modified  object, and adopted technique.}
\begin{tabular}{cccc}
\hline
 & \begin{tabular}[c]{@{}c@{}}Adversarial\\ Attack\end{tabular} & \begin{tabular}[c]{@{}c@{}}Backdoor\\ Attack\end{tabular} & \begin{tabular}[c]{@{}c@{}}Weight\\ Attack\end{tabular} \\ \hline
\multirow{2}{*}{Stage} & \multirow{2}{*}{Inference} & \multirow{2}{*}{Training} & \multirow{2}{*}{Deployment} \\
 &  &  &  \\
\begin{tabular}[c]{@{}c@{}}Modified\\ Object\end{tabular} & Test Sample & \begin{tabular}[c]{@{}c@{}}Training\\ Data\end{tabular} & \begin{tabular}[c]{@{}c@{}}Model\\ Parameters\end{tabular} \\
\multirow{2}{*}{Technique} & \multirow{2}{*}{\begin{tabular}[c]{@{}c@{}}Imperceptible\\ Noise\end{tabular}} & \multirow{2}{*}{Trigger} & \multirow{2}{*}{Bit-flip} \\
 &  &  &  \\ \hline
\end{tabular}
\label{tab:attack_compare}
\end{table}

\subsection{Adversarial Attack}
Adversarial attack modifies samples in the inference stage by adding small perturbations that remain imperceptible to the human vision system \cite{akhtar2018threat}. 
Since adversarial attack only modifies inputs while keeping the model unchanged, it has no effect on the benign samples. Besides the basic white-box attack, the black-box attack is regarded as a greater threat, including transfer-based methods \cite{wu2020skip,chen2020boosting} and query-based methods \cite{zhao2019design,andriushchenko2020square}. Many works \cite{moosavi2017universal,mopuri2018generalizable} have also shown the existence of the image-agnostic universal perturbations. Inspired by its success in the classification, it also has been extended to other tasks, including image captioning \cite{xu2019exact}, retrieval \cite{bai2020targeted}, person re-identification \cite{bai2020adversarial},  $etc.$ To mitigate the potential threat of the adversarial attack, recent studies have demonstrated many defense methods, including the preprocessing-based defense \cite{xie2017mitigating}, the detection-based defense \cite{xu2017feature}, and the adversarial learning-based defense \cite{CarmonRSDL19}.

\subsection{Backdoor Attack}
Backdoor attack happens in the training stage and requires that the attacker can tamper with the training data even the training process \cite{liu2020survey,li2020backdoor}. 
Through poisoning some training samples with a trigger, the attacker can control the behavior of the attacked DNN in the inference stage. 
For example, images with reflections are misclassified into a target class, while benign images are classified normally \cite{liu2020reflection}. Recent clean-label attacks \cite{turner2019label,zhao2020clean} improve previous methods by only poisoning training examples while leaving their labels unchanged. Besides, instead of using the fixed trigger, dynamic backdoor attacks  \cite{nguyen2020inputaware,nguyen2021wanet} generate the trigger varying from input to input, which makes the attacked models more stealthy.
Many defense methods against backdoor attack have been proposed, such as the preprocessing-based defense \cite{liu2017neural}, the model reconstruction-based defense \cite{liu2018fine}, and the trigger synthesis-based defense \cite{wang2019neural}.

\subsection{Weight Attack}
Weight attack \cite{liu2017fault,LiuMALZW018,hong2019terminal} modifies model parameters in the deployment stage, which is the studied paradigm in this work. 
It received extensive attention, since the robust model weight is the cornerstone of employing DNNs in many security-critical applications \cite{weng2020towards}. Firstly, two schemes are proposed in \cite{liu2017fault} to modify model parameters for misclassification without and with considering stealthiness, which is dubbed single bias attack (SBA) and gradient descent attack (GDA) respectively.
After that, Trojan attack \cite{LiuMALZW018} and model-reuse attack \cite{ji2018model} were proposed, which inject malicious behavior to the DNN by retraining the model. These methods require to change lots of parameters.
Recently, fault sneaking attack (FSA) \cite{zhao2019fault} was proposed, which aims to misclassify certain samples into a target class by modifying the DNN parameters with two constraints, including maintaining the classification accuracy of other samples and minimizing parameter modifications. 

\textbf{Bit-Flip based Attack.} 
Recently, some physical fault injection techniques \cite{agoyan2010flip,kim2014flipping,selmke2015precise} were proposed, which can be adopted to precisely flip any bit in the memory. Those techniques promote researchers to study how to modify model parameters at the bit-level. 
As a branch of weight attack, the bit-flip based attack was firstly explored in \cite{rakin2019bit}. It proposed an untargeted attack that can convert the attacked DNN to a random output generator with several bit-flips.
Besides, the targeted bit Trojan (TBT) \cite{rakin2020tbt} injects the fault into DNNs by flipping some critical bits. Specifically, the attacker flips the identified bits to force the network to classify all samples embedded with a trigger to a certain target class, while the network operates with normal inference accuracy with benign samples. 
Most recently, the targeted bit-flip attack (T-BFA) \cite{rakin2020t} achieves malicious purposes without modifying samples. 
Specifically, T-BFA can mislead samples from single source class or all classes to a target class by flipping the identified weight bits.
It is worth noting that the above bit-flip based attacks leverage heuristic strategies to identify critical weight bits. How to find critical bits for the bit-flip based attack method is still an important open question.

To mitigate the affect of the weight attack, many works have investigated the defense mechanisms. Previous studies \cite{rakin2020t,he2020defending} observed that increasing the network  capacity can improve the robustness against the bit-flip based attack. Moreover, the training strategies in \cite{he2020defending,stutz2021bit,stutz2021random,sun2021exploring} can improve the robustness of the model parameters.
There exist other works considering defense in the inference stage, such as weight reconstruction-based defense \cite{li2020defending}, detection-based defense \cite{he2019sensitive,li2021radar}, and Error Correction Codes (ECC)-based defense \cite{guan2019place,huang2020functional}.
For a more comprehensive comparison, we also evaluate the resistance of attack methods to defense strategies in \cite{rakin2020t,he2020defending}.

\section{Methodology}
\label{meth}

\subsection{Preliminaries}
\label{sec:prel}



\textbf{Storage and Calculation of Quantized DNNs.} 
Currently, it is a widely-used technique to quantize DNNs before deploying on devices for efficiency and reducing storage size.
For each weight in $l$-th layer of a $Q$-bit quantized DNN, it will be represented and then stored as the signed integer in two’s complement representation ($\bm{v}=[v_Q; v_{Q-1}; ...; v_1] \in \{0,1\}^{Q}$) in the memory. Attacker can modify the weights of DNNs through flipping the stored binary bits.
In this work, we adopt the layer-wise uniform weight quantization scheme similar to Tensor-RT \cite{migacz20178}. Accordingly, each binary vector $\bm{v}$ can be converted to a real number by a function $h(\cdot)$, as follow: 
\begin{equation}
h(\bm{v})=(-2^{Q-1} \cdot v_Q + \sum_{i=1}^{Q-1}{2^{i-1} \cdot v_i}) \cdot \Delta^l,
\label{eq:h}
\end{equation}
where $l$ indicates which layer the weight is from, $\Delta^l >0$ is a known and stored constant which represents the step size of the $l$-th layer weight quantizer.

\textbf{Notations.}
We denote a $Q$-bit quantized DNN-based classification model as $f:\mathcal{X} \rightarrow \mathcal{Y}$, where $\mathcal{X} \in \mathbb{R}^{d}$ being the input space and  $\mathcal{Y} \in \{1, 2, ..., K\}$ being the $K$-class output space. Assuming that the last layer of this DNN model is a fully-connected layer with $\mathbf{B} \in \{0,1\}^{K \times C \times Q}$ being the quantized weights, where $C$ is the dimension of last layer's input. Let $\mathbf{B}_{i,j} \in \{0, 1\}^{Q}$ be the two's complement representation of a single weight and $\mathbf{B}_i \in \{0,1\}^{C \times Q}$ denotes all the binary weights connected to the $i$-th output neuron. 
Given a test sample $\bm{x}_i$ with the ground-truth label $y_i$, $f(\bm{x}_i;\bm{\Theta}, \mathbf{B}) \in [0,1]^{K}$ is the output probability vector and $g(\bm{x}_i;\bm{\Theta}) \in \mathbb{R}^{C}$ is the input of the last layer, where $\bm{\Theta}$ denotes the model parameters except the last layer.

\textbf{Attack Scenario.} In this paper, we focus on the white-box bit-flip based weight attack, which was first introduced in \cite{rakin2019bit}. Specifically, we assume that the attacker has full knowledge of the model (including it's architecture, parameters, and parameters’ location in the memory), and can precisely flip any bit in the memory. Besides, we also assume that attackers can have access to a small portion of benign samples, but they can not tamper the training process and the training data. We consider two basic goals for our attack: \textit{effectiveness} and \textit{stealthiness}. The effectiveness corresponds to a higher attack success rate, while stealthiness requires reducing the effect on samples except the attacked sample(s). We consider the targeted attack with the purpose of misclassifying sample(s) to a target class $t$, since the targeted attack is more challenging than the untargeted attack. 

\subsection{General Formulation}
\label{sec:genfor}
In this section, we give the general formulation of the proposed weight attack. As mentioned in Section \ref{sec:prel}, we assume that the attacker can get access to two sample sets to perform attack: $\bm{D}_1=\{(\bm{x}_i,y_i)\}_{i=1}^{N_1}$ contributes to achieve the attack effectiveness and $\bm{D}_2=\{(\bm{x}_i,y_i)\}_{i=1}^{N_2}$ helps to keep the attack stealthiness. Let the vectorized binary parameters $\bm{b} \in \{0,1\}^V$ being a part of the attacked model ($e.g.$, the parameters of the last layer $\mathbf{B}$) which the attacker intends to modify, and $\hat{\bm{b}} \in \{0,1\}^V$ corresponds to the modified version of $\bm{b}$. 
The dimension $V$ will be specified later. 
When performing attack, we only modify $\bm{b}$ and the remaining parameters of the attacked model are fixed. Note that we will omit the fixed parameters of the attacked model for clarity in the below formulas. Let $\phi$ with the learnable parameter $\bm{q}$ represent the modification operator on $\bm{D}_1$, which is supposed to be differentiable $w.r.t.$ $\bm{q}$. The overall objective function is below:
\begin{equation}
\begin{split}
    \underset{\hat{\bm{b}}, \bm{q}}{\mathrm{min}} & \quad \lambda_1 \mathcal{L}_1(\phi(\bm{D}_1;\bm{q}), t;\hat{\bm{b}}) + \lambda_2 \mathcal{L}_2(\bm{D}_2;\hat{\bm{b}}), \quad
    \\ &  \mathrm{s.t.} \ \ \hat{\bm{b}} \in \{0,1\} ^ {V}, \ \ d_H(\bm{b}, \hat{\bm{b}}) \leq k,
\end{split}
\label{eq:overall}
\end{equation}
where $d_H(\cdot, \cdot)$ denotes the Hamming distance, and $\lambda_1, \lambda_2 > 0$ are the trade-off parameters. The loss $\mathcal{L}_1$ is used to ensure the attack effectiveness, which can be customized according to the attacker's purpose. By minimizing $\mathcal{L}_1$, the attacked model can misclassify the attacked sample(s) ($e.g.$, a specified sample or samples with a trigger) into a target class $t$, as shown in Fig. \ref{fig:demo}. We will detail the loss $\mathcal{L}_1$ later for two weight attack versions, respectively. 

The attack may be easily detectable using only $\mathcal{L}_1$, since the attacked model behaves abnormally even on samples except the attacked sample(s). Therefore, we utilize the loss $\mathcal{L}_2$ to ensure the attack stealthiness. The loss $\mathcal{L}_2$ is formulated as follows, 
\begin{equation}
\begin{split}
    \mathcal{L}_{2}(\bm{D}_2;\hat{\bm{b}})=\sum_{(\bm{x}_i,y_i) \in \bm{D}_2}{\ell(f(\bm{x}_i;\hat{\bm{b}}), y_i)},
\end{split}
\label{eq:genl2}
\end{equation}
where $f_j(\bm{x}_i;\hat{\bm{b}})$ indicates the posterior probability of $\bm{x}_i$ $w.r.t.$ class $j$. $\ell(\cdot, \cdot)$ is specified by the cross entropy loss.

Besides, to better meet our goal, a straightforward additional approach is reducing the magnitude of the modification. In this paper, we constrain the number of bit-flips less than $k$. Physical bit flipping techniques can be time-consuming as discussed in \cite{van2016drammer,zhao2019fault}. Moreover, such techniques lead to abnormal behaviors in the attacked system ($e.g.$, suspicious cache activity of processes), which may be detected by some physical detection-based defenses \cite{gruss2018another}. As such, limiting the number of bit-flips is critical to make the attack more efficient and practical. 

\subsection{Single Sample Attack}
In this section, based on the above general formulation, we introduce our first type of attack: single sample attack (SSA). 

\textbf{Attacker's Goals.} For SSA, the goals of \textit{effectiveness} and \textit{stealthiness} can be described as follows. The effectiveness requires that the attacked model can misclassify a specific sample to a predefined target class without any sample modification, and the stealthiness requires that the prediction accuracy of other samples will not be significantly reduced. SSA aims at misclassifying a specific sample, whose goal may correspond to the attacker's requirement in some scenarios. For example, the attacker wants to manipulate the behavior of the intrusion detection system on a specific input. 
Moreover, as the minimal malicious requirement, misclassifying a single sample may result in a low number of bit-flips, which will be demonstrated in our experiments. 

\textbf{Loss for Ensuring Effectiveness.} 
Recall that SSA aims at forcing a specific image to be classified as the target class by modifying the model parameters at the bit-level. To this end, the most straightforward way is maximizing the logit of the target class while minimizing that of the source class. For a sample $\bm{x}$ with the ground-truth label $s$, the logit of a class can be directly determined by the input of the last layer $g(\bm{x};\bm{\Theta})$ and weights connected to the node of that class. Accordingly, we can modify weights only connected to the source and target class to fulfill our purpose. Moreover, for SSA, we specify the set $\bm{D}_1$ as $\{(\bm{x}, s)\}$ and use $\bm{x}$ and $s$ straightly in the below formulations for clarity. We have no modification on the attacked sample, $i.e.$, $\phi$ in Eq. (\ref{eq:overall}) is specified as $\phi(\{(\bm{x},s)\}, \bm{q})=\{(\bm{x},s)\}$ and will be omitted below. The loss for ensuring effectiveness is as follows:
\begin{equation}
\begin{split}
        \mathcal{L}_1^{SSA}(\bm{x}, s, t; \hat{\mathbf{B}}_s, \hat{\mathbf{B}}_t) & =
        \max\big(\tau-p(\bm{x}; \bm{\Theta}, \hat{\mathbf{B}}_t)+\delta, 0\big)   \\ & + \max\big(p(\bm{x}; \bm{\Theta}, \hat{\mathbf{B}}_s)-\tau + \delta, 0\big),
\end{split}
\end{equation}
where $p(\bm{x}; \bm{\Theta}, \hat{\mathbf{B}}_i)=[h(\hat{\mathbf{B}}_{i,1});h(\hat{\mathbf{B}}_{i,2});...;h(\hat{\mathbf{B}}_{i,C})]^{\top}g(\bm{x};\bm{\Theta})$ denotes the logit of class $i$ ($i=s$ or $i=t$),   
$h(\cdot)$ is the function defined in Eq. (\ref{eq:h}), 
$\tau=\underset{i \in \{0,...,K\}\backslash{\{s\}}}{\max}p(\bm{x}; \bm{\Theta}, \mathbf{B}_i)$, and $\delta \in \mathbb{R}$ indicates a slack variable, which will be specified in later experiments. 
The first term of $\mathcal{L}_1^{SSA}$ aims at increasing the logit of the target class, while the second term is to decrease the logit of the source class. The loss $\mathcal{L}_1^{SSA}$ is 0 only when the output on target class is more than $\tau+\delta$ and the output on source class is less than $\tau-\delta$. That is, the prediction on $\bm{x}$ of the target model is the predefined target class $t$.
%


\textbf{Loss for Ensuring Stealthiness.} For SSA, we use $\mathcal{L}_2$  to ensure the stealthiness. $\mathcal{L}_2$ defined by Eq. (\ref{eq:genl2}) can be rewritten for SSA using  $\hat{\mathbf{B}}_s$ and $\hat{\mathbf{B}}_t$, as follows:
\begin{equation}
\begin{split}
         \mathcal{L}_2(\bm{D}_2;\hat{\mathbf{B}}_s, \hat{\mathbf{B}}_t)=\sum_{(\bm{x}_i,y_i)\in\bm{D}_2}{\ell(f(\bm{x}_i; \hat{\mathbf{B}}_s, \hat{\mathbf{B}}_t), y_i)}.
\end{split}
\end{equation}

\textbf{Overall Objective for SSA.} In conclusion, the final objective function for SSA is as follows:
\begin{equation}
\begin{split}
    \underset{\hat{\mathbf{B}}_s, \hat{\mathbf{B}}_t}{\mathrm{min}} \quad & \lambda_1 \mathcal{L}_1^{SSA}(\bm{x}, s, t;\hat{\mathbf{B}}_s, \hat{\mathbf{B}}_t) + \lambda_2 \mathcal{L}_2(\bm{D}_2;\hat{\mathbf{B}}_s, \hat{\mathbf{B}}_t), \quad
    \\ \mathrm{s.t.} &  \ \ \hat{\mathbf{B}}_{s} \in \{0,1\} ^ {C \times Q},  \ \ \hat{\mathbf{B}}_{t} \in \{0,1\} ^ {C \times Q}, \\ & d_H(\mathbf{B}_s, \hat{\mathbf{B}}_{s}) + d_H(\mathbf{B}_t, \hat{\mathbf{B}}_{t}) \leq k.
\end{split}
\label{eq:ssa_obj}
\end{equation}
Note that the above objective function is one of special forms of Eq. (\ref{eq:overall}), where $\phi$ is the identity function. $\hat{\mathbf{B}}_s, \hat{\mathbf{B}}_t \in \{0,1\}^{C \times Q}$ are two variables we want to optimize, corresponding to the weights of the fully-connected layer $w.r.t.$ class $s$ and $t$, respectively, in the attacked DNN model.
$\mathbf{B} \in \{0,1\}^{K\times C \times Q}$ denotes the weights of the fully-connected layer of the original DNN model. 
Therefore, $\bm{b}$ in Section \ref{sec:genfor} corresponds to the reshaped and concatenated $\mathbf{B}_s$ and $\mathbf{B}_t$ and $\hat{\bm{b}}$ could be the reshaped and concatenated $\hat{\mathbf{B}}_s$ and $\hat{\mathbf{B}}_t$, respectively. The size of $\bm{b}$ and $\hat{\bm{b}}$ is $2CQ$ ($i.e.$, $V=2CQ$).

\subsection{Triggered Samples Attack}

In this section, we present an attack considering embedding the inputs with a learned trigger, namely, triggered samples attack (TSA).


\textbf{Attacker's Goals.} 
For TSA, the \textit{effectiveness} goal means that the attacked model can classify the samples embedded with a designed trigger ($e.g.$, a square patch) to a target class, and the \textit{stealthiness} goal expects that the attacked model performs accurate classification on most inputs when the trigger is removed.
Compared to SSA, TSA can mislead the attacked model on any samples with the trigger, which may be suitable to the scenarios that require various malicious inputs. For example, in the context of autonomous driving, the attacker wishes that autonomous cars recognize all road signs with the trigger as the stop sign. Besides, the normal behavior of the attacked model when the trigger is absent makes the attack undetectable.

\textbf{Loss for Ensuring Effectiveness.} Since TSA utilizes the trigger to perform attack, we can suppose that two sample sets $\bm{D}_1$ and $\bm{D}_2$ share the same data, $i.e.$,  $\bm{D}=\bm{D}_1=\bm{D}_2$ and $N=N_1=N_2$. We use $\bm{D}$ for clarity below. To achieve the effectiveness goal, we present how to embed the inputs with a trigger firstly. We suppose that the trigger is a patch and its area is given by an attacker-specified mask $\bm{m} \in [0,1]^d$. We define the function $\phi$ to generate the triggered samples, as follows.
\begin{equation}
\begin{split}
    \phi(\bm{D};\bm{q})=\{((1-\bm{m}) \otimes  \bm{x}_i + \bm{m} \otimes \bm{q}, y_i) \mid (\bm{x}_i,y_i) \in \bm{D} \},
\end{split}
\label{eq:adding_trigger}
\end{equation}
where $\otimes$ indicates the element-wise product and $\bm{q} \in \mathbb{R}^d$ is the trigger. Because the modification on the sample is differentiable $w.r.t.$ $\bm{q}$, the trigger can be optimized using the gradient method to achieve a more powerful attack. Note that TSA aims at misclassifying the samples with the trigger from all classes, which is different from SSA. Therefore, we optimize the parameters of the last layer $\mathbf{B}$ and $\hat{\mathbf{B}}$ is the modified $\mathbf{B}$. Based on the above modification, we define the following objective to achieve the targeted misclassification on the triggered sample.
\begin{equation}
\begin{split}
    & \mathcal{L}_{1}^{TSA}(\phi(\bm{D};\bm{q}), t; \hat{\mathbf{B}})\\ = & \sum_{(\bm{x}_i,y_i)\in\bm{D}}{\ell(f((1-\bm{m}) \otimes  \bm{x}_i + \bm{m} \otimes \bm{q}; \hat{\mathbf{B}}), t)},
\end{split}
\end{equation}
where $\ell(\cdot)$ is the cross entropy loss. 
We can update the trigger $\bm{q}$ and $\hat{\mathbf{B}}$ alternatively to find the trigger and the bit-flips to minimize the above loss.

\textbf{Loss for Ensuring Stealthiness.} Since we modify $\mathbf{B}$ for TSA, $\mathcal{L}_2$ defined by Eq. (\ref{eq:genl2}) can be rewritten as below.

\begin{equation}
\begin{split}
    \mathcal{L}_{2}(\bm{D};\hat{\mathbf{B}})=\sum_{(\bm{x}_i,y_i)\in\bm{D}}{\ell(f(\bm{x}_i;\hat{\mathbf{B}}), y_i)}.
\end{split}
\end{equation}

\textbf{Overall Objective for TSA.} 
We summarize the objective function of TSA as below.

\begin{equation}
\begin{split}
    \underset{\hat{\mathbf{B}}, \bm{q}}{\mathrm{min}} \quad & \lambda_1 \mathcal{L}_1^{TSA}(\phi(\bm{D};\bm{q}), t;\hat{\mathbf{B}}) + \lambda_2 \mathcal{L}_2(\bm{D};\hat{\mathbf{B}}) 
    \\  \mathrm{s.t.} \ \ \hat{\mathbf{B}} & \in \{0,1\} ^ {K \times C \times Q}, \ \ d_H(\mathbf{B}, \hat{\mathbf{B}}) \leq k.
\end{split}
\label{eq:tsa_obj}
\end{equation}
Note that we specify $\bm{q}$ as the learnable trigger pattern and the modification operator $\phi$ as embedding this trigger (see Eq. (\ref{eq:adding_trigger})). Since the weights of the fully-connected layer $\hat{\mathbf{B}} \in \{0,1\}^{K\times C \times Q}$ is the variable we want to optimize, $\bm{b}$ in Section \ref{sec:genfor} corresponds to the reshaped $\mathbf{B}$ and $\hat{\bm{b}}$ could be the reshaped $\hat{\mathbf{B}}$. The size of $\bm{b}$ and $\hat{\bm{b}}$ is $KCQ$ ($i.e.$, $V=KCQ$).

\subsection{An Effective Optimization Method}

To solve the challenging MIP problem (\ref{eq:overall}), we adopt the generic solver for integer programming, dubbed $\ell_p$-Box ADMM \cite{wu2018ell}. The solver presents its superior performance in many tasks, $e.g.$, model pruning \cite{li2019compressing}, clustering \cite{bibi2019constrained}, MAP inference \cite{wu2020map}, adversarial attack \cite{fan2020sparse}, $etc.$ It proposed to replace the binary constraint equivalently by the intersection of two continuous constraints, as follows
\begin{equation}
    \hat{\bm{b}} \in \{0,1\} ^ {V} \Leftrightarrow \hat{\bm{b}} \in (\mathcal{S}_b \cap \mathcal{S}_p),
\label{eq:cons}
\end{equation}
where $\mathcal{S}_b =[0,1]^{V}$ indicates the box constraint, and $\mathcal{S}_p=\{\hat{\bm{b}}: || \hat{\bm{b}}-\frac{\boldsymbol{1}}{2} ||_2^{2}=\frac{V}{4}\}$ denotes the $\ell_2$-sphere constraint.
Utilizing (\ref{eq:cons}), Problem (\ref{eq:overall}) can be equivalently reformulated. Besides, for binary vector $\bm{b}$ and $\hat{\bm{b}}$, there exists a nice relationship between Hamming distance and Euclidean distance: $d_H(\bm{b}, \hat{\bm{b}})=||\bm{b}-\hat{\bm{b}}||_2^2$. The reformulated objective is as follows:
\begin{equation}
\begin{split}
     \underset{\hat{\bm{b}},\bm{q}, \bm{u}_1 \in \mathcal{S}_b,\bm{u}_2\in \mathcal{S}_p, u_3 \in \mathbb{R}^+ }{\mathrm{min}} & \quad \lambda_1 \mathcal{L}_1(\phi(\bm{D}_1;\bm{q});\hat{\bm{b}}) + \lambda_2 \mathcal{L}_2(\bm{D}_2;\hat{\bm{b}}), \\
      \mathrm{s.t.} ~~ \hat{\bm{b}}=\bm{u}_1, & \hat{\bm{b}}=\bm{u}_2,  ||  \bm{b}-\hat{\bm{b}}||_2^2 - k + u_3 = 0, 
\end{split}
\end{equation}
where two extra variables $\bm{u}_1, \bm{u}_2 \in \mathbb{R}^{V}$ are introduced to split the constraints $w.r.t.$ $\hat{\bm{b}}$. 
Besides, the non-negative slack variable $u_3\in \mathbb{R}^+$ is used to transform $||  \bm{b}-\hat{\bm{b}}||_2^2 - k \leq 0$ in (\ref{eq:overall}) into $||  \bm{b}-\hat{\bm{b}}||_2^2 - k + u_3= 0$.
The above constrained optimization problem can be efficiently solved by the alternating direction method of multipliers (ADMM) \cite{boyd2011distributed}. 

Following the standard procedure of ADMM, we firstly present the augmented Lagrangian function of the above problem, as follows:
\begin{equation}
\begin{split}
     & L(\hat{\bm{b}}, \bm{q}, \bm{u}_1,  \bm{u}_2,  u_3, \bm{z}_1,  \bm{z}_2, z_3 )  \\ =  & \lambda_1 \mathcal{L}_1(\phi(\bm{D}_1;\bm{q});\hat{\bm{b}}) + \lambda_2 \mathcal{L}_2(\bm{D}_2;\hat{\bm{b}})  \\
      + & \bm{z}_1^\top(\hat{\bm{b}}-\bm{u}_1)  +   
     \bm{z}_2^\top(\hat{\bm{b}}-\bm{u}_2) 
     + z_3 (||  \bm{b}-\hat{\bm{b}}||_2^2 - k + u_3) \\ 
      + & c_1(\bm{u}_1)  +  c_2(\bm{u}_2)    +  c_3(u_3) \\ 
      +  &  \frac{\rho_1}{2}||  \hat{\bm{b}} - \bm{u}_1||_2^2 + \frac{\rho_2}{2}||  \hat{\bm{b}} - \bm{u}_2||_2^2 + \frac{\rho_3}{2}( ||  \bm{b}-\hat{\bm{b}}||_2^2 - k + u_3 )^2,
\end{split}
\label{eq: augmented lagrangian}
\end{equation}
where $\bm{z}_1, \bm{z}_2 \in \mathbb{R}^{V}$ and $z_3 \in \mathbb{R}$ are dual variables, and $\rho_1$, $\rho_2$, $\rho_3 > 0$ are penalty factors, which will be specified later. $c_1(\bm{u}_1)=\mathbb{I}_{ \{\bm{u}_1 \in \mathcal{S}_b\} }$, $c_2(\bm{u}_2)=\mathbb{I}_{ \{\bm{u}_2 \in \mathcal{S}_p\} }$, and $c_3(u_3)=\mathbb{I}_{ \{u_3 \in \mathbb{R}^+\} }$ capture the constraints $\mathcal{S}_b, \mathcal{S}_p$ and $\mathbb{R}^+$, respectively. The indicator function $\mathbb{I}_{\{a\}}=0$ if $a$ is true; otherwise, $\mathbb{I}_{\{a\}}=+\infty$. 
Based on the augmented Lagrangian function, the primary and dual variables are updated iteratively, with $r$ indicating the iteration index.

\textbf{Given $(\hat{\bm{b}}^r, \bm{q}^{r}, \bm{z}_1^r, \bm{z}_2^r, z_3^r)$, update $(\bm{u}_1^{r+1}, \bm{u}_2^{r+1}, u_3^{r+1})$.} 
Given $(\hat{\bm{b}}^r, \bm{q}, \bm{z}_1^r, \bm{z}_2^r, z_3^r)$, $(\bm{u}_1,\bm{u}_2, u_3)$ are independent, and they can be optimized in parallel, as follows:
\begin{equation}
\begin{split}
\left\{\begin{array}{l}
\bm{u}_1^{r+1} =\underset{\bm{u}_1 \in \mathcal{S}_b}{\text{arg min}} \ \ (\bm{z}_1^r)^\top(\hat{\bm{b}}^r-\bm{u}_1) + \frac{\rho_1}{2}||  \hat{\bm{b}}^r - \bm{u}_1||_2^2 \\ ~~~~~~~~ = \mathcal{P}_{\mathcal{S}_b}(\hat{\bm{b}}^r +\frac{\bm{z}_1^r}{\rho_1}), 
\\
\bm{u}_2^{r+1}=\underset{\bm{u}_2 \in \mathcal{S}_p}{\text{arg min}} \ \ (\bm{z}_2^r)^\top(\hat{\bm{b}}^r-\bm{u}_2) + \frac{\rho_2}{2}||  \hat{\bm{b}}^r - \bm{u}_2||_2^2 \\ ~~~~~~~~ = \mathcal{P}_{\mathcal{S}_p}(\hat{\bm{b}}^r+\frac{\bm{z}_2^r}{\rho_2}),
\\
u_3^{r+1}=\underset{u_3 \in \mathbb{R}^+}{\text{arg min}} \ \   z_3^r (||  \bm{b}-\hat{\bm{b}}^r||_2^2 - k + u_3) \\ ~~~~~~~~~~~~~~~~~~~~~~~ + \frac{\rho_3}{2}( ||  \bm{b}-\hat{\bm{b}}^r||_2^2 - k + u_3 )^2 
\\
~~~~~~~~ = \mathcal{P}_{\mathbb{R}^+}(-||  \bm{b}-\hat{\bm{b}}^r||_2^2 + k - \frac{z_3^r}{\rho_3} ),
\end{array}\right. 
\end{split}    
\label{eq:subpro}
\end{equation}
where $\mathcal{P}_{\mathcal{S}_b}(\bm{a}) \! =\! \mathrm{min}((\bm{1}, \max(\bm{0}, \bm{a}))$ with $ \bm{a}  \! \in \! \mathbb{R}^n$ is the projection onto the box constraint $\mathcal{S}_b$; 
$\mathcal{P}_{\mathcal{S}_p}(\bm{a})=\frac{\sqrt{n}}{2} \frac{\bar{\bm{a}}}{|| \bm{a} ||} + \frac{\bm{1}}{2}$ with $\bar{\bm{a}}= \bm{a}-\frac{\bm{1}}{2}$ indicates the projection onto the $\ell_2$-sphere constraint $\mathcal{S}_p$ \cite{wu2018ell};
$\mathcal{P}_{\mathbb{R}^+}(a) \! = \! \max(0, a)$ with $a \! \in \! \mathbb{R}$ indicates the projection onto $\mathbb{R}^+$.

\textbf{Given $(\bm{u}_1^{r+1}, \! \bm{u}_2^{r+1}, \! u_3^{r+1}, \!  \bm{z}_1^r, \! \bm{z}_2^r, \! z_3^r)$, update $\hat{\bm{b}}^{r+1}$ and $\bm{q}^{r}$.}
Although there is no closed-form solution to $\hat{\bm{b}}^{r+1}$, it can be easily updated by the gradient descent method, as $\mathcal{L}_1(\phi(\bm{D}_1;\bm{q}), t;\hat{\bm{b}})$
and $\mathcal{L}_2(\bm{D}_2;\hat{\bm{b}})$ are differentiable \textit{\textit{w.r.t.}} $\hat{\bm{b}}$, as follows
\begin{equation}
\hat{\bm{b}}^{r+1} \leftarrow \hat{\bm{b}}^r - \eta \cdot \frac{\partial L(\hat{\bm{b}}, \bm{q}^r, \bm{u}_1^{r+1}, \bm{u}_2^{r+1}, u_3^{r+1}, \bm{z}_1^r, \bm{z}_2^r, z_3^r)}{\partial \hat{\bm{b}}}\Big|_{\hat{\bm{b}}=\hat{\bm{b}}^r},
\label{eq: update b}
\end{equation}
where $\eta > 0$ denotes the step size. Note that we can run multiple steps of gradient descent in the above update. Both the number of steps and $\eta$ will be specified in later experiments. 
Besides, the detailed derivation of $\partial L/ \partial \hat{\bm{b}}$ for SSA and TSA can be found in \textbf{Appendix}.

As mentioned in Section \ref{sec:genfor}, we suppose that $\phi$ is differentiable $w.r.t.$ $\bm{q}$ and thus 
$\mathcal{L}_1(\phi(\bm{D}_1;\bm{q});\hat{\bm{b}})$ is differentiable $w.r.t.$ $\bm{q}$. We can update $\bm{q}$ using the gradient descent method as follows
\begin{equation}
\bm{q}^{r+1} \leftarrow \bm{q}^r - \zeta \cdot \frac{\partial L(\hat{\bm{b}}^{r}, \bm{q}, \bm{u}_1^{r+1}, \bm{u}_2^{r+1}, u_3^{r+1}, \bm{z}_1^r, \bm{z}_2^r, z_3^r)}{\partial \bm{q}}\Big|_{\bm{q}=\bm{q}^r},
\label{eq: update q}
\end{equation}
where $\zeta > 0$ denotes the step size. The update of $\bm{q}$  is kept pace with $\hat{\bm{b}}$, $i.e.$, we update $\bm{q}$ for one step for each update of $\hat{\bm{b}}$. Note that the update of $\bm{q}$ is kept pace with $\hat{\bm{b}}$, using same backward pass.

\begin{algorithm}[t]
\caption{Continuous optimization for the proposed bit-flip based weight attack} 
{\bf Input:} 
The original quantized DNN model $f$ with weights $\bm{\Theta}, \mathbf{B}$; specified $\mathcal{L}_1$, $\mathcal{L}_2$, and $\phi$; target class $t$; auxiliary sample set $\bm{D}_1$ and $\bm{D}_2$; hyper-parameters $\lambda_1$, $\lambda_2$, and $k$.\\
{\bf Output:}  
$\hat{\bm{b}}$.
\begin{algorithmic}[1]
\State  Initial $\bm{u}_1^0$, $\bm{u}_2^0$, $u_3^0$, $\bm{z}_1^0$, $\bm{z}_2^0$, $z_3^0$, $\hat{\bm{b}}^0$, $\bm{q}^0$ and let $r \leftarrow 0$;
\While {not converged}
    \State Update $\bm{u}_1^{r+1}$, $\bm{u}_2^{r+1}$ and $u_3^{r+1}$ as Eq. (\ref{eq:subpro});
    \State Update $\hat{\bm{b}}^{r+1}$ as Eq. (\ref{eq: update b}); 
    \State Update $\hat{\bm{q}}^{r+1}$ as Eq. (\ref{eq: update q}); 
    \State Update $\bm{z}_1^{r+1}$, $\bm{z}_2^{r+1}$ and $z_3^{r+1}$ as Eq. (\ref{eq:z});
    \State $r \leftarrow r+1$.
\EndWhile
\end{algorithmic}
\label{alg:framework}
\end{algorithm}

\textbf{Given $(\hat{\bm{b}}^{r+1}, \bm{q}^{r+1}, \bm{u}_1^{r+1}, \bm{u}_2^{r+1}, u_3^{r+1})$,  update $(\bm{z}_1^{r+1}, $ $\bm{z}_2^{r+1}, z_3^{r+1})$.} 
The dual variables are updated by the gradient ascent method, as follows
\begin{equation}
\begin{split}
\left\{\begin{array}{l}
    \bm{z}_1^{r+1} = \bm{z}_1^r + \rho_1(\hat{\bm{b}}^{r+1}-\bm{u}_1^{r+1}), \\
    \bm{z}_2^{r+1} = \bm{z}_2^r + \rho_2(\hat{\bm{b}}^{r+1}-\bm{u}_2^{r+1}), \\
    z_3^{r+1} = z_3^r + \rho_3 (||  \bm{b}-\hat{\bm{b}}^{r+1}||_2^2 - k + u_3^{r+1}).
\end{array}\right.
\end{split}
\label{eq:z}
\end{equation}

\textbf{Remarks.} 
{\bf 1)} Note that since the dual variables $(\bm{u}_1^{r+1}, \bm{u}_2^{r+1}, u_3^{r+1})$ are updated in parallel, and the primal variables $(\hat{\bm{b}}, \bm{q})$ are also updated in parallel, the above algorithm is a two-block ADMM algorithm. We summarize this algorithm in Algorithm \ref{alg:framework}.
{\bf 2)} Except for the update of $\hat{\bm{b}}^{r+1}$ and $\bm{q}^{r+1}$, other updates are very simple and efficient. The computational cost of the whole algorithm will be analyzed in Section \ref{sec:complexity}.
{\bf 3)} Due to the inexact solution to $\hat{\bm{b}}^{r+1}$ and $\bm{q}^{r+1}$ using gradient descent, the theoretical convergence of the whole ADMM algorithm cannot be guaranteed.
However, as demonstrated in many previous works \cite{gol1979modified,eckstein1992douglas,boyd2011distributed}, the inexact two-block ADMM often shows good practical convergence, which is also the case in our later experiments. Besides, the numerical convergence analysis is presented in Section \ref{sec:convergence}.
{\bf 4)} The proper adjustment of $(\rho_1, \rho_2, \rho_3)$ could accelerate the practical convergence, which will be specified later.

\section{Experiments}

\subsection{Evaluation Setup}
\subsubsection{Datasets}
We conduct experiments on CIFAR-10 \cite{krizhevsky2009learning} and ImageNet \cite{russakovsky2015imagenet}. 
CIFAR-10 consists of 50K training samples and 10K testing samples with 10 classes. ImageNet contains 1.2M samples from the training set and 50K samples from the validation set, categorized into 1,000 classes. 
For the methods in Section \ref{sec:ssa_exp} , we randomly select 1,000 images from each dataset as the evaluation set. 
Specifically, for each of the 10 classes in CIFAR-10, we perform attacks on the 100 randomly selected validation images from the other 9 classes. 
For ImageNet, we randomly choose 50 target classes.  
For each target class, we perform attacks on 20 images randomly selected from the rest classes in the validation set.  
For the methods in Section \ref{sec:tsa_exp}, we select all 10 classes for CIFAR-10 and randomly select 5 classes for ImageNet as the target classes. For each target class, we use all testing samples (10K images for CIFAR-10 and 50K images for ImageNet) with the generated trigger to evaluate the attack performance.  

Besides, for all methods in Section \ref{sec:ssa_exp} and \ref{sec:tsa_exp} except GDA which does not employ auxiliary samples, we provide 128 and 512 auxiliary samples on CIFAR-10 and ImageNet, respectively, which corresponds to the size of $\bm{D}_2$ ($N_2$) for SSA and the size of $\bm{D}$ ($N$) for TSA. 

\begin{table*}[t]
\caption{Results of five attack methods across different bit-widths and architectures on CIFAR-10 and ImageNet (bold: the best; underline: the second best). The mean and standard deviation of PA-ACC and $\mathrm{N_{flip}}$ are calculated by attacking the 1,000 images. Our method is denoted as \textbf{SSA}.}
\label{tab:main}
\centering
\resizebox{180mm}{40mm}{
\begin{tabular}{cccccccccc}
\hline
\multicolumn{1}{c|}{Dataset} & \multicolumn{1}{c|}{Method} & \multicolumn{1}{c|}{\begin{tabular}[c]{@{}c@{}}Target \\ Model\end{tabular}} & \begin{tabular}[c]{@{}c@{}}PA-ACC\\ ($\%$)\end{tabular} & \begin{tabular}[c]{@{}c@{}}ASR\\ ($\%$)\end{tabular} & \multicolumn{1}{c|}{$\mathrm{N_{flip}}$} & \multicolumn{1}{c|}{\begin{tabular}[c]{@{}c@{}}Target \\ Model\end{tabular}} & \begin{tabular}[c]{@{}c@{}}PA-ACC\\ ($\%$)\end{tabular} & \begin{tabular}[c]{@{}c@{}}ASR\\ ($\%$)\end{tabular} & $\mathrm{N_{flip}}$ \\ \hline
\specialrule{0em}{0pt}{-6pt}\\ \hline
\multicolumn{1}{c|}{\multirow{12}{*}{\rotatebox{90}{CIFAR-10}}} & \multicolumn{1}{c|}{FT} & \multicolumn{1}{c|}{\multirow{5}{*}{\begin{tabular}[c]{@{}c@{}}ResNet 8-bit\\ \\ ACC: 92.16$\%$\end{tabular}}} & 85.01\scriptsize   {$\pm$2.90} & \textbf{100.0} & \multicolumn{1}{c|}{1507.51\scriptsize   {$\pm$86.54}} & \multicolumn{1}{c|}{\multirow{5}{*}{\begin{tabular}[c]{@{}c@{}}VGG 8-bit\\ \\ ACC: 93.20$\%$\end{tabular}}} & 84.31\scriptsize   {$\pm$3.10} & 98.7 & 11298.74\scriptsize   {$\pm$830.36} \\
\multicolumn{1}{c|}{} & \multicolumn{1}{c|}{T-BFA} & \multicolumn{1}{c|}{} & 87.56\scriptsize {$\pm$2.22} & 98.7 & \multicolumn{1}{c|}{\underline{9.91\scriptsize   {$\pm$2.33}}} & \multicolumn{1}{c|}{} & \textbf{89.83\scriptsize {$\pm$3.92}} & 96.7 & \underline{14.53\scriptsize   {$\pm$3.74}} \\
\multicolumn{1}{c|}{} & \multicolumn{1}{c|}{FSA} & \multicolumn{1}{c|}{} & \textbf{88.38\scriptsize {$\pm$2.28}} & 98.9 & \multicolumn{1}{c|}{185.51\scriptsize   {$\pm$54.93}} & \multicolumn{1}{c|}{} & \underline{88.80\scriptsize {$\pm$2.86}} & 96.8 & 253.92\scriptsize   {$\pm$122.06} \\
\multicolumn{1}{c|}{} & \multicolumn{1}{c|}{GDA} & \multicolumn{1}{c|}{} & 86.73\scriptsize {$\pm$3.50} & 99.8 & \multicolumn{1}{c|}{26.83\scriptsize   {$\pm$12.50}} & \multicolumn{1}{c|}{} & 85.51\scriptsize {$\pm$2.88} & \textbf{100.0} & 21.54\scriptsize   {$\pm$6.79} \\
\multicolumn{1}{c|}{} & \multicolumn{1}{c|}{\textbf{SSA}} & \multicolumn{1}{c|}{} & \underline{88.20\scriptsize {$\pm$2.64}} & \textbf{100.0} & \multicolumn{1}{c|}{\textbf{5.57\scriptsize   {$\pm$1.58}}} & \multicolumn{1}{c|}{} & 86.06\scriptsize {$\pm$3.17} & \textbf{100.0} & \textbf{7.40\scriptsize   {$\pm$2.72}} \\ \cline{2-10} 
\multicolumn{1}{c|}{} & \multicolumn{1}{c|}{FT} & \multicolumn{1}{c|}{\multirow{5}{*}{\begin{tabular}[c]{@{}c@{}}ResNet 4-bit\\ \\ ACC: 91.90$\%$\end{tabular}}} & 84.37\scriptsize {$\pm$2.94} & \textbf{100.0} & \multicolumn{1}{c|}{392.48\scriptsize   {$\pm$47.26}} & \multicolumn{1}{c|}{\multirow{5}{*}{\begin{tabular}[c]{@{}c@{}}VGG 4-bit\\ \\ ACC: 92.61$\%$\end{tabular}}} & 83.31\scriptsize {$\pm$3.76} & 94.5 & 2270.52\scriptsize   {$\pm$324.69} \\
\multicolumn{1}{c|}{} & \multicolumn{1}{c|}{T-BFA} & \multicolumn{1}{c|}{} & 86.46\scriptsize {$\pm$2.80} & 97.9 & \multicolumn{1}{c|}{\underline{8.80\scriptsize   {$\pm$2.01}}} & \multicolumn{1}{c|}{} & \textbf{88.74\scriptsize {$\pm$4.52}} & 96.2 & 11.23\scriptsize   {$\pm$2.36} \\
\multicolumn{1}{c|}{} & \multicolumn{1}{c|}{FSA} & \multicolumn{1}{c|}{} & \underline{87.73\scriptsize {$\pm$2.36}} & 98.4 & \multicolumn{1}{c|}{76.83\scriptsize   {$\pm$25.27}} & \multicolumn{1}{c|}{} & \underline{87.58\scriptsize {$\pm$3.06}} & 97.5 & 75.03\scriptsize   {$\pm$29.75} \\
\multicolumn{1}{c|}{} & \multicolumn{1}{c|}{GDA} & \multicolumn{1}{c|}{} & 86.25\scriptsize {$\pm$3.59} & 99.8 & \multicolumn{1}{c|}{14.08\scriptsize   {$\pm$7.94}} & \multicolumn{1}{c|}{} & 85.08\scriptsize {$\pm$2.82} & \textbf{100.0} & \underline{10.31\scriptsize   {$\pm$3.77}} \\
\multicolumn{1}{c|}{} & \multicolumn{1}{c|}{\textbf{SSA}} & \multicolumn{1}{c|}{} & \textbf{87.82\scriptsize {$\pm$2.60}} & \textbf{100.0} & \multicolumn{1}{c|}{\textbf{5.25\scriptsize   {$\pm$1.09}}} & \multicolumn{1}{c|}{} & 85.91\scriptsize {$\pm$3.29} & \textbf{100.0} & \textbf{6.26\scriptsize   {$\pm$2.37}} \\ \hline
\specialrule{0em}{0pt}{-6pt}\\ \hline
\multicolumn{1}{c|}{\multirow{12}{*}{\rotatebox{90}{ImageNet}}} & \multicolumn{1}{c|}{FT} & \multicolumn{1}{c|}{\multirow{5}{*}{\begin{tabular}[c]{@{}c@{}}ResNet 8-bit\\ \\ ACC: 69.50$\%$\end{tabular}}} & 59.33\scriptsize   {$\pm$0.93} & \textbf{100.0} & \multicolumn{1}{c|}{277424.29\scriptsize   {$\pm$12136.34}} & \multicolumn{1}{c|}{\multirow{5}{*}{\begin{tabular}[c]{@{}c@{}}VGG 8-bit\\ \\ ACC: 73.31$\%$\end{tabular}}} & 62.08\scriptsize   {$\pm$2.33} & \textbf{100.0} & 1729685.22\scriptsize   {$\pm$137539.54} \\
\multicolumn{1}{c|}{} & \multicolumn{1}{c|}{T-BFA} & \multicolumn{1}{c|}{} & 68.71\scriptsize {$\pm$0.36} & 79.3 & \multicolumn{1}{c|}{24.57\scriptsize   {$\pm$20.03}} & \multicolumn{1}{c|}{} & 73.09\scriptsize {$\pm$0.12} & 84.5 & 363.78\scriptsize   {$\pm$153.28} \\
\multicolumn{1}{c|}{} & \multicolumn{1}{c|}{FSA} & \multicolumn{1}{c|}{} & \underline{69.27\scriptsize {$\pm$0.15}} & 99.7 & \multicolumn{1}{c|}{441.21\scriptsize   {$\pm$119.45}} & \multicolumn{1}{c|}{} & \underline{73.28\scriptsize {$\pm$0.03}} & \textbf{100.0} & 1030.03\scriptsize   {$\pm$260.30} \\
\multicolumn{1}{c|}{} & \multicolumn{1}{c|}{GDA} & \multicolumn{1}{c|}{} & 69.26\scriptsize {$\pm$0.22} & \textbf{100.0} & \multicolumn{1}{c|}{\underline{18.54\scriptsize   {$\pm$6.14}}} & \multicolumn{1}{c|}{} & \textbf{73.29\scriptsize {$\pm$0.02}} & \textbf{100.0} & \underline{197.05\scriptsize   {$\pm$49.85}} \\
\multicolumn{1}{c|}{} & \multicolumn{1}{c|}{\textbf{SSA}} & \multicolumn{1}{c|}{} & \textbf{69.41\scriptsize   {$\pm$0.08}} & \textbf{100.0} & \multicolumn{1}{c|}{\textbf{7.37\scriptsize   {$\pm$2.18}}} & \multicolumn{1}{c|}{} & \underline{73.28\scriptsize   {$\pm$0.03}} & \textbf{100.0} & \textbf{69.89\scriptsize   {$\pm$18.42}} \\ \cline{2-10} 
\multicolumn{1}{c|}{} & \multicolumn{1}{c|}{FT} & \multicolumn{1}{c|}{\multirow{5}{*}{\begin{tabular}[c]{@{}c@{}}ResNet 4-bit\\ \\ ACC: 66.77$\%$\end{tabular}}} & 15.65\scriptsize   {$\pm$4.52} & \textbf{100.0} & \multicolumn{1}{c|}{135854.50\scriptsize   {$\pm$21399.94}} & \multicolumn{1}{c|}{\multirow{5}{*}{\begin{tabular}[c]{@{}c@{}}VGG  4-bit\\ \\ ACC: 71.76$\%$\end{tabular}}} & 17.76\scriptsize   {$\pm$1.71} & \textbf{100.0} & 1900751.70\scriptsize   {$\pm$37329.44} \\
\multicolumn{1}{c|}{} & \multicolumn{1}{c|}{T-BFA} & \multicolumn{1}{c|}{} & 65.86\scriptsize {$\pm$0.42} & 80.4 & \multicolumn{1}{c|}{24.79\scriptsize   {$\pm$19.02}} & \multicolumn{1}{c|}{} & 71.49\scriptsize {$\pm$0.15} & 84.3 & 350.33\scriptsize   {$\pm$158.57} \\
\multicolumn{1}{c|}{} & \multicolumn{1}{c|}{FSA} & \multicolumn{1}{c|}{} & 66.44\scriptsize {$\pm$0.21} & 99.9 & \multicolumn{1}{c|}{157.53\scriptsize   {$\pm$33.66}} & \multicolumn{1}{c|}{} & 71.69\scriptsize {$\pm$0.09} & \textbf{100.0} & 441.32\scriptsize   {$\pm$111.26} \\
\multicolumn{1}{c|}{} & \multicolumn{1}{c|}{GDA} & \multicolumn{1}{c|}{} & \underline{66.54\scriptsize {$\pm$0.22}} & \textbf{100.0} & \multicolumn{1}{c|}{\underline{11.45\scriptsize   {$\pm$3.82}}} & \multicolumn{1}{c|}{} & \textbf{71.73\scriptsize {$\pm$0.03}} & \textbf{100.0} & \underline{107.18\scriptsize   {$\pm$28.70}} \\
\multicolumn{1}{c|}{} & \multicolumn{1}{c|}{\textbf{SSA}} & \multicolumn{1}{c|}{} & \textbf{66.69\scriptsize   {$\pm$0.07}} & \textbf{100.0} & \multicolumn{1}{c|}{\textbf{7.96\scriptsize   {$\pm$2.50}}} & \multicolumn{1}{c|}{} & \textbf{71.73\scriptsize   {$\pm$0.03}} & \textbf{100.0} & \textbf{69.72\scriptsize   {$\pm$18.84}} \\ \hline
\end{tabular}}
\label{tab:main_ssa}
\end{table*}

\subsubsection{Target Models}
According to the setting in \cite{rakin2020tbt,rakin2020t}, we adopt two popular network architectures: ResNet \cite{he2016deep} and VGG \cite{simonyan2014very} for evaluation.
On CIFAR-10, we perform experiments on ResNet-20 and VGG-16. 
On ImageNet, we use the pre-trained ResNet-18\footnote{Downloaded from \url{https://download.pytorch.org/models/resnet18-5c106cde.pth}} and VGG-16\footnote{Downloaded from \url{https://download.pytorch.org/models/vgg16_bn-6c64b313.pth}} network. 
We quantize all networks to the 4-bit and 8-bit quantization level using the layer-wise uniform weight quantization scheme, which is similar to the one involved in the Tensor-RT solution \cite{migacz20178}. 

\subsubsection{Evaluation Metrics}
\label{sec:eval_metric}
We adopt three metrics to evaluate the attack performance, \textit{i.e.,} the post attack accuracy (PA-ACC), the attack success rate (ASR), and the number of bit-flips ($\mathrm{N_{flip}}$). 
PA-ACC denotes the post attack accuracy on the original validation set. 
ASR is defined as the ratio of attacked samples that are successfully attacked into the target class. To be specific, we calculate ASR using all 1,000 attacked samples for the methods in Section \ref{sec:ssa_exp} and all testing samples with the trigger for the methods in Section \ref{sec:tsa_exp}. Therefore, we can calculate an ASR after 1,000 attacks for the methods in Section \ref{sec:ssa_exp} and obtain an ASR after an attack for the methods in Section \ref{sec:tsa_exp}. 
$\mathrm{N_{flip}}$ is the number of bit-flips required for an attack. 
A better attack performance corresponds to a higher PA-ACC and ASR, while a lower $\mathrm{N_{flip}}$. 
Besides, we also show the accuracy of the original model, denoted as ACC.

\subsubsection{Defense Methods}
Besides attacking the standard training models, we also test the resistance of all attacks to two defense methods: piece-wise clustering \cite{he2020defending} and larger model capacity \cite{he2020defending,rakin2020t}.

He $et$ $al.$ \cite{he2020defending} proposed a novel training technique, called piece-wise clustering, to enhance the network robustness against the bit-flip based attack. 
Such a training technique introduces an additional weight penalty to the inference loss, which has the effect of eliminating close-to-zero weights \cite{he2020defending}. 
We test the resistance of all attack methods to the piece-wise clustering. 
We conduct experiments with the 8-bit quantized ResNet on CIFAR-10 and ImageNet. Following the ideal configuration in \cite{he2020defending}, the clustering coefficient, which is a hyper-parameter of piece-wise clustering, is set to 0.001 in our evaluation. 

Previous studies \cite{he2020defending,rakin2020t} observed that increasing the network capacity can improve the robustness against the bit-flip based attack. 
Accordingly, we evaluate all attack methods against the models with a larger capacity using the 8-bit quantized ResNet on both datasets. 
Similar to the strategy in \cite{he2020defending}, we increase the model capacity by varying the network width ($i.e.$, 2$\times$ width in our experiments).

We conduct experiments with the 8-bit quantized ResNet on CIFAR-10 and ImageNet to evaluate the attack performance of all attack methods against these two defense methods. Besides the three metrics in Section \ref{sec:eval_metric}, we also present the number of increased $\mathrm{N_{flip}}$ compared to the model without defense ($i.e.$, results in Table \ref{tab:main_ssa} and \ref{tab:main_tsa}), denoted as $\Delta \mathrm{N_{flip}}$.

\begin{table*}[]
\caption{Results of all attack methods against the models with defense on CIFAR-10 and ImageNet (bold: the best; underline: the second best). The mean and standard deviation of PA-ACC and $\mathrm{N_{flip}}$ are calculated by attacking the 1,000 images. Our method is denoted as \textbf{SSA}. $\Delta \mathrm{N_{flip}}$ denotes the increased $\mathrm{N_{flip}}$ compared to the corresponding result in Table \ref{tab:main}.}
\label{tab:defense}
\centering
\resizebox{150mm}{40mm}{

\setlength{\tabcolsep}{4mm}{
\begin{tabular}{cccccccc}
\hline
\multicolumn{1}{l|}{Defense} & \multicolumn{1}{c|}{Dataset} & \multicolumn{1}{c|}{Method} & \multicolumn{1}{c|}{\begin{tabular}[c]{@{}c@{}}ACC\\ ($\%$)\end{tabular}} & \begin{tabular}[c]{@{}c@{}}PA-ACC\\ ($\%$)\end{tabular} & \begin{tabular}[c]{@{}c@{}}ASR\\ ($\%$)\end{tabular} & $\mathrm{N_{flip}}$ & $\Delta   \mathrm{N_{flip}}$ \\ \hline
\specialrule{0em}{0pt}{-6pt}\\ \hline
\multicolumn{1}{c|}{\multirow{10}{*}{\rotatebox{90}{Piece-wise Clustering}}} & \multicolumn{1}{c|}{\multirow{5}{*}{CIFAR-10}} & \multicolumn{1}{c|}{FT} & \multicolumn{1}{c|}{\multirow{6}{*}{91.01}} & 84.06\scriptsize   {$\pm$3.56} & 99.5 & 1893.55\scriptsize   {$\pm$68.98} & 386.04 \\
\multicolumn{1}{c|}{} & \multicolumn{1}{c|}{} & \multicolumn{1}{c|}{T-BFA} & \multicolumn{1}{c|}{} & 85.82\scriptsize {$\pm$1.89} & 98.6 & \underline{45.51\scriptsize   {$\pm$9.47}} & 35.60 \\
\multicolumn{1}{c|}{} & \multicolumn{1}{c|}{} & \multicolumn{1}{c|}{FSA} & \multicolumn{1}{c|}{} & \underline{86.61\scriptsize {$\pm$2.51}} & 98.6 & 246.11\scriptsize   {$\pm$75.36} & 60.60 \\
\multicolumn{1}{c|}{} & \multicolumn{1}{c|}{} & \multicolumn{1}{c|}{GDA} & \multicolumn{1}{c|}{} & 84.12\scriptsize {$\pm$4.77} & \textbf{100.0} & 52.76\scriptsize   {$\pm$16.29} & \underline{25.93} \\
\multicolumn{1}{c|}{} & \multicolumn{1}{c|}{} & \multicolumn{1}{c|}{\textbf{SSA}} & \multicolumn{1}{c|}{} & \textbf{87.30\scriptsize {$\pm$2.74}} & \textbf{100.0} & \textbf{18.93\scriptsize   {$\pm$7.11}} & \textbf{13.36} \\ \cline{2-8} 
\multicolumn{1}{c|}{} & \multicolumn{1}{c|}{\multirow{5}{*}{ImageNet}} & \multicolumn{1}{c|}{FT} & \multicolumn{1}{c|}{\multirow{6}{*}{63.62}} & 43.44\scriptsize {$\pm$2.07} & 92.2 & 762267.56\scriptsize   {$\pm$52179.46} & 484843.27 \\
\multicolumn{1}{c|}{} & \multicolumn{1}{c|}{} & \multicolumn{1}{c|}{T-BFA} & \multicolumn{1}{c|}{} & 62.82\scriptsize {$\pm$0.27} & 90.1 & 273.56\scriptsize   {$\pm$191.29} & 248.99 \\
\multicolumn{1}{c|}{} & \multicolumn{1}{c|}{} & \multicolumn{1}{c|}{FSA} & \multicolumn{1}{c|}{} & \underline{63.26\scriptsize {$\pm$0.21}} & 99.5 & 729.94\scriptsize   {$\pm$491.83} & 288.73 \\
\multicolumn{1}{c|}{} & \multicolumn{1}{c|}{} & \multicolumn{1}{c|}{GDA} & \multicolumn{1}{c|}{} & 63.14\scriptsize {$\pm$0.48} & \textbf{100.0} & \underline{107.59\scriptsize   {$\pm$31.15}} & \underline{89.05} \\
\multicolumn{1}{c|}{} & \multicolumn{1}{c|}{} & \multicolumn{1}{c|}{\textbf{SSA}} & \multicolumn{1}{c|}{} & \textbf{63.52\scriptsize   {$\pm$0.14}} & \textbf{100.0} & \textbf{51.11\scriptsize   {$\pm$4.33}} & \textbf{43.74} \\ \hline
\specialrule{0em}{0pt}{-6pt}\\ \hline
\multicolumn{1}{c|}{\multirow{10}{*}{\rotatebox{90}{Larger Model Capacity}}} & \multicolumn{1}{c|}{\multirow{5}{*}{CIFAR-10}} & \multicolumn{1}{c|}{FT} & \multicolumn{1}{c|}{\multirow{6}{*}{94.29}} & 86.46\scriptsize   {$\pm$2.84} & \textbf{100.0} & 2753.43\scriptsize   {$\pm$188.27} & 1245.92 \\
\multicolumn{1}{c|}{} & \multicolumn{1}{c|}{} & \multicolumn{1}{c|}{T-BFA} & \multicolumn{1}{c|}{} & \textbf{91.16\scriptsize {$\pm$1.42}} & 98.7 & \underline{17.91\scriptsize   {$\pm$4.64}} & \underline{8.00} \\
\multicolumn{1}{c|}{} & \multicolumn{1}{c|}{} & \multicolumn{1}{c|}{FSA} & \multicolumn{1}{c|}{} & 90.70\scriptsize {$\pm$2.37} & 98.5 & 271.27\scriptsize   {$\pm$65.18} & 85.76 \\
\multicolumn{1}{c|}{} & \multicolumn{1}{c|}{} & \multicolumn{1}{c|}{GDA} & \multicolumn{1}{c|}{} & 89.83\scriptsize {$\pm$3.02} & \textbf{100.0} & 48.96\scriptsize   {$\pm$21.03} & 22.13 \\
\multicolumn{1}{c|}{} & \multicolumn{1}{c|}{} & \multicolumn{1}{c|}{\textbf{SSA}} & \multicolumn{1}{c|}{} & \underline{90.96\scriptsize {$\pm$2.63}} & \textbf{100.0} & \textbf{8.79\scriptsize   {$\pm$2.44}} & \textbf{3.22} \\ \cline{2-8} 
\multicolumn{1}{c|}{} & \multicolumn{1}{c|}{\multirow{5}{*}{ImageNet}} & \multicolumn{1}{c|}{FT} & \multicolumn{1}{c|}{\multirow{6}{*}{71.35}} & 63.51\scriptsize {$\pm$1.29} & \textbf{100.0} & 507456.61\scriptsize   {$\pm$34517.04} & 230032.32 \\
\multicolumn{1}{c|}{} & \multicolumn{1}{c|}{} & \multicolumn{1}{c|}{T-BFA} & \multicolumn{1}{c|}{} & 70.84\scriptsize {$\pm$0.30} & 88.9 & 40.23\scriptsize   {$\pm$27.29} & 15.66 \\
\multicolumn{1}{c|}{} & \multicolumn{1}{c|}{} & \multicolumn{1}{c|}{FSA} & \multicolumn{1}{c|}{} & \textbf{71.30\scriptsize {$\pm$0.04}} & \textbf{100.0} & 449.70\scriptsize   {$\pm$106.42} & 8.49 \\
\multicolumn{1}{c|}{} & \multicolumn{1}{c|}{} & \multicolumn{1}{c|}{GDA} & \multicolumn{1}{c|}{} & \textbf{71.30\scriptsize {$\pm$0.05}} & \textbf{100.0} & \underline{20.01\scriptsize   {$\pm$6.04}} & \underline{1.47} \\
\multicolumn{1}{c|}{} & \multicolumn{1}{c|}{} & \multicolumn{1}{c|}{\textbf{SSA}} & \multicolumn{1}{c|}{} & \textbf{71.30\scriptsize   {$\pm$0.04}} & \textbf{100.0} & \textbf{8.48\scriptsize   {$\pm$2.52}} & \textbf{1.11} \\ \hline
\end{tabular}}}
\label{tab:defense_ssa}
\end{table*}

\subsection{Results of SSA}
\label{sec:ssa_exp}

\subsubsection{Baseline Methods}
We compare our SSA with GDA \cite{liu2017fault}, FSA \cite{zhao2019fault}, and T-BFA \cite{rakin2020t}. Since GDA \cite{liu2017fault} and FSA \cite{zhao2019fault} are originally designed for attacking the full-precision network, we adapt these two methods to attack the quantized network by applying quantization-aware training \cite{jacob2018quantization}.
We adopt the $\ell_0$-norm for FSA \cite{liu2017fault} and modification compression for GDA \cite{zhao2019fault} to reduce the number of the modified parameters. 
Among three types of T-BFA \cite{rakin2020t}, we compare to the most comparable method: the 1-to-1 stealthy attack scheme. 
The purpose of this attack scheme is to misclassify samples of a single source class into the target class while maintaining the prediction accuracy of other samples.
Besides, we take the fine-tuning (FT) of the last fully-connected layer using the objective defined in Eq. (\ref{eq:ssa_obj}) without considering the constraints as a basic attack and present its results. 

\subsubsection{Implementation Details of SSA}
\label{sec:set_ssa}
For each attack, we fix $\lambda_1$ as 1 and adopt a strategy for jointly searching $\lambda_2$ and $k$. 
Specifically, for an initially given $k$, we search $\lambda_2$ from a relatively large initial value and divide it by 2 if the attack does not succeed.
The maximum search times of $\lambda_2$ for a fixed $k$ is set to 8. 
If it exceeds the maximum search times, we double $k$ and search $\lambda_2$ from the relatively large initial value. 
The maximum search times of $k$ is set to 4. 
On CIFAR-10, the initial $k$ and $\lambda_2$ are set to 5 and 100. 
On ImageNet, $\lambda_2$ is initialized as $10^4$; 
$k$ is initialized as 5 and 50 for ResNet and VGG, respectively. 
On CIFAR-10, the $\delta$ in $\mathcal{L}_1$ is set to 10. 
On ImageNet, the $\delta$ is set to 3 and increased to 10 if the attack fails. 
$\bm{u}_1$ and $\bm{u}_2$ are initialized as $\bm{b}$ and $u_3$ is initialized as 0. 
$\bm{z}_1$ and $\bm{z}_2$ are initialized as $\bm{0}$ and $z_3$ is initialized as 0. 
$\hat{\bm{b}}$ is initialized as $\bm{b}$. 
During each iteration, the number of gradient steps for updating $\hat{\bm{b}}$ is 5 and the step size is set to 0.01 on both datasets.
Hyper-parameters ($\rho_1$, $\rho_2$, $\rho_3$) (see Eq. (\ref{eq:z})) are initialized as ($10^{-4}$, $10^{-4}$, $10^{-5}$) on both datasets, and increased by $\rho_i \leftarrow \rho_i \times 1.01$, $i=1,2,3$ after each iteration. 
The maximum values of ($\rho_1$, $\rho_2$, $\rho_3$) are set to (50, 50, 5) on both datasets. 
Besides the maximum number of iterations ($i.e.$, 2,000), we also set another stopping criterion, $i.e.$, $||  \hat{\bm{b}}- \bm{u}_1||_2^2 \leq 10^{-4}$ and $|| \hat{\bm{b}}- \bm{u}_2||_2^2 \leq 10^{-4}$.

\subsubsection{Main Results}
\textbf{Results on CIFAR-10.}
The results of all methods on CIFAR-10 are shown in Table \ref{tab:main_ssa}. 
Our method achieves a 100\% ASR with the fewest $\mathrm{N_{flip}}$ for all the bit-widths and architectures. 
FT modifies the maximum number of bits among all methods since there is no limitation of parameter modifications. 
Due to the absence of the training data, the PA-ACC of FT is also poor. 
These results indicate that fine-tuning the trained DNN as an attack method is infeasible. 
Although T-BFA flips the second-fewest bits under three cases, it fails to achieve a higher ASR than GDA and FSA. 
In terms of PA-ACC, SSA is comparable to other methods. 
Note that the PA-ACC of SSA significantly outperforms that of GDA, which is the most competitive \textit{w.r.t.} ASR and $\mathrm{N_{flip}}$ among all the baseline methods. 
The PA-ACC of GDA is relatively poor, because it does not employ auxiliary samples. 
Achieving the highest ASR, the lowest $\mathrm{N_{flip}}$, and the comparable PA-ACC demonstrates that our optimization-based method is more superior to other heuristic methods (T-BFA and GDA).

\textbf{Results on ImageNet.}
The results on ImageNet are shown in Table \ref{tab:main_ssa}. 
It can be observed that GDA shows very competitive performance compared to other methods. 
However, our method obtains the highest PA-ACC, the fewest bit-flips (less than 8), and a 100\% ASR in attacking ResNet. 
For VGG, our method also achieves a 100\% ASR with the fewest $\mathrm{N_{flip}}$ for both bit-widths.  
The $\mathrm{N_{flip}}$ results of our method are mainly attributed to the cardinality constraint on the number of bit-flips. 
Moreover, for our method, the average PA-ACC degradation over four cases on ImageNet is only 0.06\%, which demonstrates the stealthiness of our attack. 
When comparing the results of ResNet and VGG, an interesting observation is that all methods require significantly more bit-flips for VGG. 
One reason is that VGG is much wider than ResNet. 
Similar to the claim in \cite{he2020defending}, increasing the network width contributes to the robustness against the bit-flip based attack. 

\begin{figure*}[ht]
    \centering
    \includegraphics[width=0.99\textwidth]{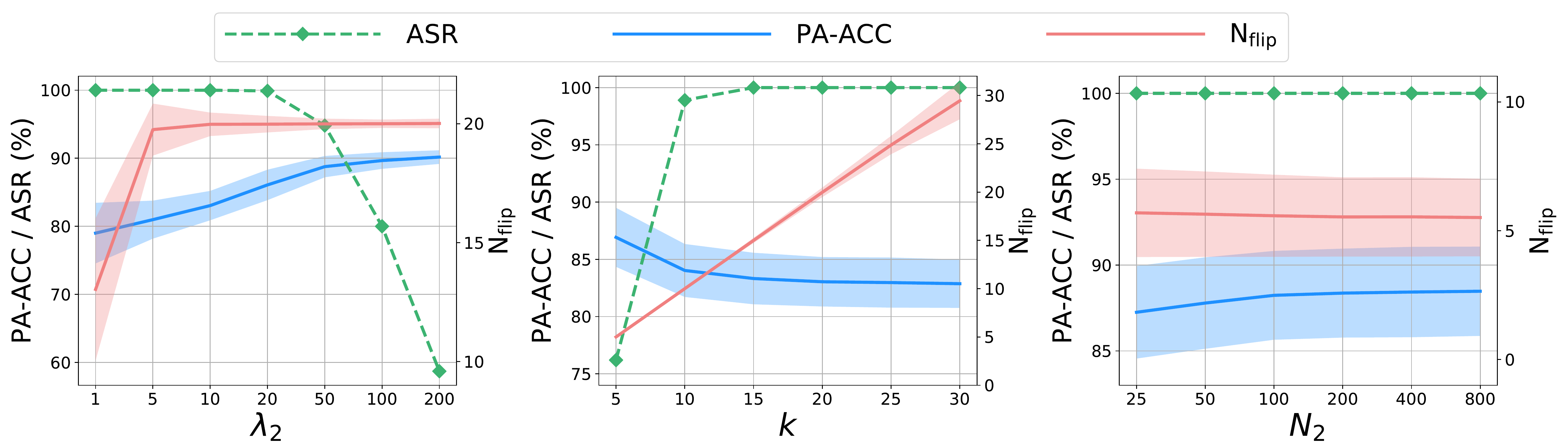}
    \caption{Results of SSA with different parameters $\lambda_2$ ($\lambda_1$ is fixed at 1), $k$, and the number of auxiliary samples $N_2$ on CIFAR-10. Regions in shadow indicate the standard deviation of attacking the 1,000 images.}
	\label{fig:ablation_study_ssa}
\end{figure*}

\begin{figure*}[h]
    \centering
    \includegraphics[width=0.99\textwidth]{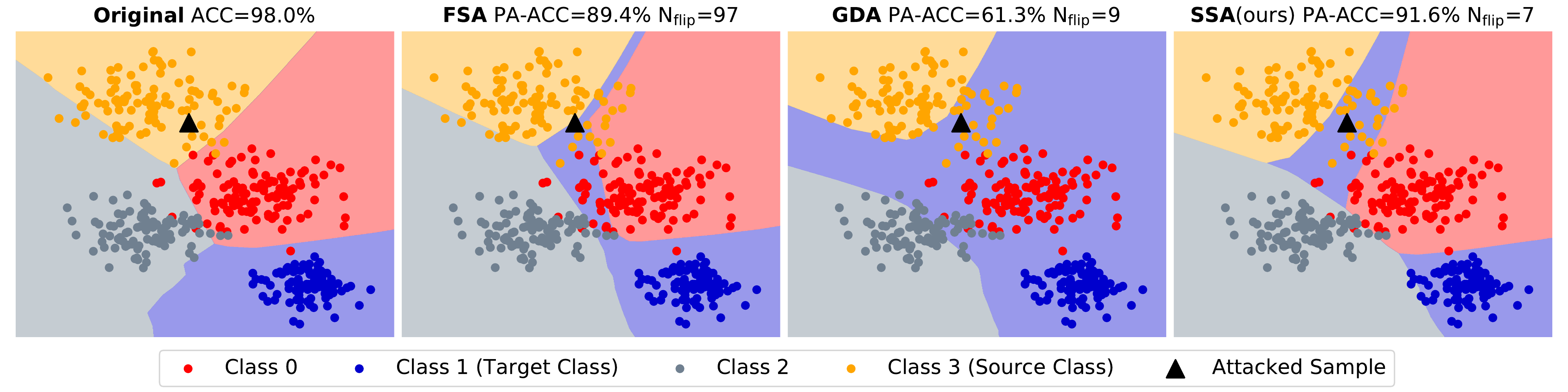}
    \caption{Visualization of decision boundaries of the original model and the post attack models. The attacked sample from Class 3 is misclassified into the Class 1 by FSA, GDA, and SSA (ours).}
	\label{fig:db}
\end{figure*}

\subsubsection{Resistance to Defense Methods}
\label{sec:defense_ssa}
\textbf{Resistance to Piece-wise Clustering.}
For our method, the initial $k$ is set to 50 on ImageNet and the rest settings are the same as those in Section \ref{sec:set_ssa}. 
The results of the resistance to the piece-wise clustering of all attack methods are shown in Table \ref{tab:defense}. 
It shows that the model trained with piece-wise clustering can improve the number of required bit-flips for all attack methods. 
However, our method still achieves a 100\% ASR with the least number of bit-flips on both two datasets. 
Compared with other methods, SSA achieves the fewest $\Delta \mathrm{N_{flip}}$ on ImageNet and the best PA-ACC on both datasets. 
These results demonstrate the superiority of our method over other methods when attacking models trained with piece-wise clustering.

\textbf{Resistance to Larger Model Capacity.}
All settings of our method are the same as those used in Section \ref{sec:set_ssa}.   
The results are presented in Table \ref{tab:defense_ssa}. 
We observe that all methods require more bit-flips to attack the model with the 2$\times$ width.
To some extent, it demonstrates that the wider network with the same architecture is more robust against the bit-flip based attack. 
However, our method still achieves a 100\% ASR with the fewest $\mathrm{N_{flip}}$ and $\Delta \mathrm{N_{flip}}$. 
Moreover, when comparing the two defense methods, we find that piece-wise clustering performs better than the model with a larger capacity in terms of $\Delta \mathrm{N_{flip}}$. 
However, piece-wise clustering training also causes the accuracy decrease of the original model  ($e.g.$, from 92.16\% to 91.01\% on CIFAR-10).

\subsubsection{Ablation Studies}
We perform ablation studies on parameters $\lambda_2$ ($\lambda_1$ is fixed at 1), $k$, and the number of auxiliary samples $N_2$. 
We use the 8-bit quantized ResNet on CIFAR-10 as the representative for analysis.
We discuss the attack performance of SSA under different values of $\lambda_2$ while $k$ is fixed at 20, and under different values of $k$ while $\lambda_2$ is fixed at 10. 
To analyze the effect of $N_2$, we configure $N_2$ from 25 to 800 and keep other settings the same as those in Section \ref{sec:set_ssa}. 
The results are presented in Fig. \ref{fig:ablation_study_ssa}. 
We observe that our method achieves a 100\% ASR when $\lambda_2$ is less than 20. 
As expected, the PA-ACC increases while the ASR decreases along with the increase of $\lambda_2$, since larger $\lambda_2$ encourages more stealthy attack. 
The plot of parameter $k$ presents that $k$ can exactly limit the number of bit-flips, while other attack methods do not involve such constraint.
This advantage is critical since it allows the attacker to identify limited bits to perform an attack when the budget is fixed. 
As shown in the figure, the number of auxiliary samples less than 200 has a marked positive impact on the PA-ACC. 
It's intuitive that more auxiliary samples can lead to a better PA-ACC. 
The observation also indicates that SSA still works well without too many auxiliary samples. 

\begin{table*}[]
\caption{Results of three attack methods across different bit-widths and architectures on CIFAR-10 and ImageNet (bold: the best). The mean and standard deviation of TA, PA-ACC, and $\mathrm{N_{flip}}$ are calculated by attacking into 10 and 5 target classes on CIFAR-10 and ImageNet, respectively. Our method is denoted as \textbf{TSA}.}
\centering
\resizebox{180mm}{27mm}{
\begin{tabular}{cccccccccc}
\hline
\multicolumn{1}{c|}{Dataset} & \multicolumn{1}{c|}{Method} & \multicolumn{1}{c|}{\begin{tabular}[c]{@{}c@{}}Target \\ Model\end{tabular}} & \begin{tabular}[c]{@{}c@{}}PA-ACC\\ ($\%$)\end{tabular} & \begin{tabular}[c]{@{}c@{}}ASR\\ ($\%$)\end{tabular} & \multicolumn{1}{c|}{$\mathrm{N_{flip}}$} & \multicolumn{1}{c|}{\begin{tabular}[c]{@{}c@{}}Target \\ Model\end{tabular}} & \begin{tabular}[c]{@{}c@{}}PA-ACC\\ ($\%$)\end{tabular} & \begin{tabular}[c]{@{}c@{}}ASR\\ ($\%$)\end{tabular} & $\mathrm{N_{flip}}$ \\
\hline
\specialrule{0em}{0pt}{-6pt}\\ \hline
\multicolumn{1}{c|}{\multirow{6}{*}{\rotatebox{90}{CIFAR-10}}} & \multicolumn{1}{c|}{FT} & \multicolumn{1}{c|}{\multirow{3}{*}{\begin{tabular}[c]{@{}c@{}}ResNet 8-bit\\ TA: 92.16$\%$\end{tabular}}} & 84.23\scriptsize{$\pm$6.67} & 84.25\scriptsize{$\pm$14.08} & \multicolumn{1}{c|}{18.9\scriptsize{$\pm$9.72}} & \multicolumn{1}{c|}{\multirow{3}{*}{\begin{tabular}[c]{@{}c@{}}VGG 8-bit\\ TA: 93.20$\%$\end{tabular}}} & 19.77\scriptsize{$\pm$10.79} & \textbf{97.97\scriptsize{$\pm$2.22}} & 2130.7\scriptsize{$\pm$829.60} \\
\multicolumn{1}{c|}{} & \multicolumn{1}{c|}{TBT} & \multicolumn{1}{c|}{} & 87.52\scriptsize{$\pm$3.69} & 85.60\scriptsize{$\pm$17.55} & \multicolumn{1}{c|}{63.7\scriptsize{$\pm$4.88}} & \multicolumn{1}{c|}{} & 78.30\scriptsize{$\pm$24.84} & 47.65\scriptsize{$\pm$24.41} & 602.7\scriptsize{$\pm$15.62} \\
\multicolumn{1}{c|}{} & \multicolumn{1}{c|}{\textbf{TSA}} & \multicolumn{1}{c|}{} & \textbf{88.09\scriptsize{$\pm$4.59}} & \textbf{96.06\scriptsize{$\pm$2.25}} & \multicolumn{1}{c|}{\textbf{4.5\scriptsize{$\pm$1.50}}} & \multicolumn{1}{c|}{} & \textbf{92.04\scriptsize{$\pm$3.04}} & 95.59\scriptsize{$\pm$1.57} & \textbf{6.6\scriptsize{$\pm$2.50}} \\ \cline{2-10} 
\multicolumn{1}{c|}{} & \multicolumn{1}{c|}{FT} & \multicolumn{1}{c|}{\multirow{3}{*}{\begin{tabular}[c]{@{}c@{}}ResNet 4-bit\\ TA: 91.90$\%$\end{tabular}}} & 82.38\scriptsize{$\pm$6.39} & 79.73\scriptsize{$\pm$18.42} & \multicolumn{1}{c|}{18.5\scriptsize{$\pm$9.88}} & \multicolumn{1}{c|}{\multirow{3}{*}{\begin{tabular}[c]{@{}c@{}}VGG 4-bit\\ TA: 92.61$\%$\end{tabular}}} & 30.75\scriptsize{$\pm$14.60} & \textbf{95.96\scriptsize{$\pm$6.35}} & 2271.1\scriptsize{$\pm$597.21} \\
\multicolumn{1}{c|}{} & \multicolumn{1}{c|}{TBT} & \multicolumn{1}{c|}{} & 86.23\scriptsize{$\pm$3.10} & 86.99\scriptsize{$\pm$11.99} & \multicolumn{1}{c|}{27.6\scriptsize{$\pm$5.48}} & \multicolumn{1}{c|}{} & 83.87\scriptsize{$\pm$2.75} & 61.72\scriptsize{$\pm$10.57} & 266.4\scriptsize{$\pm$19.71} \\
\multicolumn{1}{c|}{} & \multicolumn{1}{c|}{\textbf{TSA}} & \multicolumn{1}{c|}{} & \textbf{87.69\scriptsize{$\pm$2.74}} & \textbf{96.48\scriptsize{$\pm$1.75}} & \multicolumn{1}{c|}{\textbf{4.6\scriptsize{$\pm$0.66}}} & \multicolumn{1}{c|}{} & \textbf{89.00\scriptsize{$\pm$4.25}} & 94.22\scriptsize{$\pm$3.10} & \textbf{5.4\scriptsize{$\pm$1.91}} \\ \hline
\specialrule{0em}{0pt}{-6pt}\\ \hline
\multicolumn{1}{c|}{\multirow{6}{*}{\rotatebox{90}{ImageNet}}} & \multicolumn{1}{c|}{FT} & \multicolumn{1}{c|}{\multirow{3}{*}{\begin{tabular}[c]{@{}c@{}}ResNet 8-bit \\ TA: 69.50$\%$\end{tabular}}} & 69.24\scriptsize{$\pm$0.07} & 73.38\scriptsize{$\pm$9.43} & \multicolumn{1}{c|}{961.8\scriptsize{$\pm$293.12}} & \multicolumn{1}{c|}{\multirow{3}{*}{\begin{tabular}[c]{@{}c@{}}VGG 8-bit\\ TA: 73.31$\%$\end{tabular}}} & 71.20\scriptsize{$\pm$0.56} & 84.09\scriptsize{$\pm$3.85} & 9845.8\scriptsize{$\pm$1608.45} \\
\multicolumn{1}{c|}{} & \multicolumn{1}{c|}{TBT} & \multicolumn{1}{c|}{} & 65.99\scriptsize{$\pm$0.54} & 94.99\scriptsize{$\pm$0.86} & \multicolumn{1}{c|}{605.4\scriptsize{$\pm$18.69}} & \multicolumn{1}{c|}{} & \textbf{73.01\scriptsize{$\pm$0.05}} & 97.48\scriptsize{$\pm$2.39} & 4631.0\scriptsize{$\pm$106.84} \\
\multicolumn{1}{c|}{} & \multicolumn{1}{c|}{\textbf{TSA}} & \multicolumn{1}{c|}{} & \textbf{69.45\scriptsize{$\pm$0.02}} & \textbf{95.63\scriptsize{$\pm$2.38}} & \multicolumn{1}{c|}{\textbf{3.4\scriptsize{$\pm$1.52}}} & \multicolumn{1}{c|}{} & 72.89\scriptsize{$\pm$0.53} & \textbf{99.85\scriptsize{$\pm$0.26}} & \textbf{164.6\scriptsize{$\pm$53.80}} \\ \cline{2-10} 
\multicolumn{1}{c|}{} & \multicolumn{1}{c|}{FT} & \multicolumn{1}{c|}{\multirow{3}{*}{\begin{tabular}[c]{@{}c@{}}ResNet 4-bit\\ TA: 66.77$\%$\end{tabular}}} & \textbf{66.50\scriptsize{$\pm$0.14}} & 0.05\scriptsize{$\pm$0.10} & \multicolumn{1}{c|}{705.8\scriptsize{$\pm$492.44}} & \multicolumn{1}{c|}{\multirow{3}{*}{\begin{tabular}[c]{@{}c@{}}VGG 4-bit\\ TA: 71.76$\%$\end{tabular}}} & 68.90\scriptsize{$\pm$0.80} & 43.28\scriptsize{$\pm$8.88} & 14319.8\scriptsize{$\pm$3783.75} \\
\multicolumn{1}{c|}{} & \multicolumn{1}{c|}{TBT} & \multicolumn{1}{c|}{} & 63.35\scriptsize{$\pm$0.32} & 95.42\scriptsize{$\pm$0.43} & \multicolumn{1}{c|}{268.2\scriptsize{$\pm$11.65}} & \multicolumn{1}{c|}{} & \textbf{71.30\scriptsize{$\pm$0.02}} & 75.95\scriptsize{$\pm$42.42} & 4359.4\scriptsize{$\pm$614.54} \\
\multicolumn{1}{c|}{} & \multicolumn{1}{c|}{\textbf{TSA}} & \multicolumn{1}{c|}{} & 64.88\scriptsize{$\pm$0.81} & \textbf{99.98\scriptsize{$\pm$0.03}} & \multicolumn{1}{c|}{\textbf{11.2\scriptsize{$\pm$2.28}}} & \multicolumn{1}{c|}{} & 69.04\scriptsize{$\pm$3.38} & \textbf{99.99\scriptsize{$\pm$0.03}} & \textbf{223.8\scriptsize{$\pm$57.73}} \\ \hline
\end{tabular}}
\label{tab:main_tsa}
\end{table*}

\subsubsection{Visualization of Decision Boundary}
To further compare FSA and GDA with our method, we visualize the decision boundaries of the original and the post attack models in Fig. \ref{fig:db}. 
We adopt a four-layer Multi-Layer Perceptron trained with the simulated 2-D Blob dataset from 4 classes. 
The original decision boundary indicates that the original model classifies all data points almost perfectly. 
The attacked sample is classified into Class 3 by all methods. 
Visually, GDA modifies the decision boundary drastically, especially for Class 0. 
However, our method modifies the decision boundary mainly around the attacked sample. 
Althoug FSA is comparable to ours visually in Fig. \ref{fig:db}, it flips 10$\times$ bits than GDA and SSA. 
In terms of the numerical results, SSA achieves the best PA-ACC and the fewest $\mathrm{N_{flip}}$. 
This finding verifies that our method can achieve a successful attack even only tweaking the original classifier.

\begin{table*}[]
\centering
\caption{Results of three attack methods against the models with defense on CIFAR-10 and ImageNet (bold: the best). The mean and standard deviation of TA, PA-ACC, and $\mathrm{N_{flip}}$ are calculated by attacking into 10 and 5 target classes on CIFAR-10 and ImageNet, respectively. Our method is denoted as \textbf{TSA}. $\Delta \mathrm{N_{flip}}$ denotes the increased $\mathrm{N_{flip}}$ compared to the corresponding result in Table \ref{tab:main_tsa}}
\resizebox{130mm}{27mm}{
\begin{tabular}{lccccccc}
\hline
\multicolumn{1}{l|}{Defense} & \multicolumn{1}{c|}{Dataset} & \multicolumn{1}{c|}{Method} & \multicolumn{1}{c|}{\begin{tabular}[c]{@{}c@{}}ACC\\ ($\%$)\end{tabular}} & \begin{tabular}[c]{@{}c@{}}PA-ACC\\ ($\%$)\end{tabular} & \begin{tabular}[c]{@{}c@{}}ASR\\ ($\%$)\end{tabular} & $\mathrm{N_{flip}}$ & $\Delta   \mathrm{N_{flip}}$ \\ \hline
\specialrule{0em}{0pt}{-6pt}\\ \hline
\multicolumn{1}{l|}{\multirow{6}{*}{\begin{tabular}[c]{@{}l@{}}\rotatebox{90}{\begin{tabular}[c]{@{}l@{}}Piece-wise \\ Clustering\end{tabular}}\end{tabular}}} & \multicolumn{1}{c|}{\multirow{3}{*}{CIFAR-10}} & \multicolumn{1}{c|}{FT} & \multicolumn{1}{c|}{\multirow{3}{*}{91.01}} & \textbf{90.06\scriptsize{$\pm$1.45}} & 39.78\scriptsize{$\pm$18.39} & 14.2\scriptsize{$\pm$11.23} & \textbf{-4.7} \\
\multicolumn{1}{l|}{} & \multicolumn{1}{c|}{} & \multicolumn{1}{c|}{TBT} & \multicolumn{1}{c|}{} & 88.97\scriptsize{$\pm$2.59} & 41.27\scriptsize{$\pm$31.47} & 124.8\scriptsize{$\pm$14.42} & 61.1 \\
\multicolumn{1}{l|}{} & \multicolumn{1}{c|}{} & \multicolumn{1}{c|}{\textbf{TSA}} & \multicolumn{1}{c|}{} & 87.46\scriptsize{$\pm$5.47} & \textbf{92.08\scriptsize{$\pm$5.66}} & \textbf{9.9\scriptsize{$\pm$5.73}} & 5.4 \\ \cline{2-8} 
\multicolumn{1}{l|}{} & \multicolumn{1}{c|}{\multirow{3}{*}{ImageNet}} & \multicolumn{1}{c|}{FT} & \multicolumn{1}{c|}{\multirow{3}{*}{63.62}} & 0.10\scriptsize{$\pm$0.00} & \textbf{100.00\scriptsize{$\pm$0.00}} & 1047.6\scriptsize{$\pm$13.24} & 85.8 \\
\multicolumn{1}{l|}{} & \multicolumn{1}{c|}{} & \multicolumn{1}{c|}{TBT} & \multicolumn{1}{c|}{} & 0.10\scriptsize{$\pm$0.00} & \textbf{100.00\scriptsize{$\pm$0.00}} & 1190.6\scriptsize{$\pm$26.49} & 585.2 \\
\multicolumn{1}{l|}{} & \multicolumn{1}{c|}{} & \multicolumn{1}{c|}{\textbf{TSA}} & \multicolumn{1}{c|}{} & \textbf{56.83\scriptsize{$\pm$2.57}} & 99.94\scriptsize{$\pm$0.03} & \textbf{63.6\scriptsize{$\pm$5.73}} & \textbf{60.2} \\ \hline
\specialrule{0em}{0pt}{-6pt}\\ \hline
\multicolumn{1}{l|}{\multirow{6}{*}{\begin{tabular}[c]{@{}l@{}}\rotatebox{90}{\begin{tabular}[c]{@{}l@{}}Larger Model \\~~~ Capacity\end{tabular}}\end{tabular}}} & \multicolumn{1}{c|}{\multirow{3}{*}{CIFAR-10}} & \multicolumn{1}{c|}{FT} & \multicolumn{1}{c|}{\multirow{3}{*}{94.29}} & 69.17\scriptsize{$\pm$5.53} & 35.46\scriptsize{$\pm$23.25} & 936.4\scriptsize{$\pm$190.54} & 917.5 \\
\multicolumn{1}{l|}{} & \multicolumn{1}{c|}{} & \multicolumn{1}{c|}{TBT} & \multicolumn{1}{c|}{} & \textbf{90.22\scriptsize{$\pm$3.80}} & 62.17\scriptsize{$\pm$31.42} & 123.5\scriptsize{$\pm$10.20} & 59.8 \\
\multicolumn{1}{l|}{} & \multicolumn{1}{c|}{} & \multicolumn{1}{c|}{\textbf{TSA}} & \multicolumn{1}{c|}{} & \textbf{90.22\scriptsize{$\pm$3.56}} & \textbf{95.24\scriptsize{$\pm$3.52}} & \textbf{9.3\scriptsize{$\pm$2.72}} & \textbf{4.8} \\ \cline{2-8} 
\multicolumn{1}{l|}{} & \multicolumn{1}{c|}{\multirow{3}{*}{ImageNet}} & \multicolumn{1}{c|}{FT} & \multicolumn{1}{c|}{\multirow{3}{*}{71.35}} & 6.19\scriptsize{$\pm$4.95} & 97.53\scriptsize{$\pm$2.66} & 1824.8\scriptsize{$\pm$136.07} & 863.0 \\
\multicolumn{1}{l|}{} & \multicolumn{1}{c|}{} & \multicolumn{1}{c|}{TBT} & \multicolumn{1}{c|}{} & 66.54\scriptsize{$\pm$0.34} & 92.16\scriptsize{$\pm$0.59} & 1224.4\scriptsize{$\pm$20.92} & 619.0 \\
\multicolumn{1}{l|}{} & \multicolumn{1}{c|}{} & \multicolumn{1}{c|}{\textbf{TSA}} & \multicolumn{1}{c|}{} & \textbf{70.97\scriptsize{$\pm$0.58}} & \textbf{97.96\scriptsize{$\pm$4.01}} & \textbf{11.0\scriptsize{$\pm$7.97}} & \textbf{7.6} \\ \hline
\end{tabular}}
\label{tab:defense_tsa}
\end{table*}

\subsection{Results of TSA}
\label{sec:tsa_exp}

\subsubsection{Baseline Methods}
We compare our TSA with TBT \cite{rakin2020tbt}, which consists of two steps: trigger generation and weight bits identification. TBT has been proven to be effective, especially for flipping only several vulnerable bits \cite{rakin2020tbt}. We also present the results of fine-tuning (FT) the last fully-connected layer and optimize the trigger using the objective defined in Eq. (\ref{eq:tsa_obj}), without considering the constraints. We keep the same trigger design for FT, TBT, and TSA, $i.e.$, $40 \times 40$ for ImageNet and $10 \times 10$ for CIFAR-10 square trigger at the bottom right, as shown in Fig. \ref{fig:trigger_vis_tsa}. 

\subsubsection{Implementation Details of TSA}
\label{sec:set_tsa}

We set $\lambda_2$ as 1 and adjust $\lambda_1$ to obtain a better trade-off between effectiveness and stealthiness for TSA. On CIFAR-10, we set $\lambda_1$ as 100 and search the best $k$ from a relatively small initial value. Specifically, we start from $k=5$ and double $k$ if ASR calculated on the auxiliary sample set $\bm{D}$ is less than 98\%.  The maximum search times of $k$ is set to 4. On ImageNet, we set $\lambda_1=2\times10^4$ and $k=50$ for ResNet and $\lambda_1=
3\times10^4$ and $k=500$ for VGG. 
For both datasets, $\hat{\bm{b}}$ is initialized as $\bm{b}$ and $\bm{q}$ is initialized randomly. 
During each iteration, the number of gradient steps for updating $\hat{\bm{b}}$ and $\bm{q}$ is 5. 
The step size of updating $\hat{\bm{b}}$ is set to 0.001, and the step size of updating $\bm{q}$ is set to 1 at the first 1,000 iterations and thereafter decreased to 0.1.
The initial values of ($\bm{u}_1, \bm{u}_2, u_3$), ($\bm{z}_1, \bm{z}_2, z_3$), and ($\rho_1$, $\rho_2$, $\rho_3$) are same as the settings in Section \ref{sec:set_ssa}.
The maximum values of ($\rho_1$, $\rho_2$, $\rho_3$) are set to (100, 100, 10) on both datasets. 
Besides the maximum number of iterations ($i.e.$, 3,000), we also set another stopping criterion, $i.e.$, $||  \hat{\bm{b}}- \bm{u}_1||_2^2 \leq 10^{-6}$ and $|| \hat{\bm{b}}- \bm{u}_2||_2^2 \leq 10^{-6}$ for ResNet on ImageNet and $||  \hat{\bm{b}}- \bm{u}_1||_2^2 \leq 10^{-4}$ and $|| \hat{\bm{b}}- \bm{u}_2||_2^2 \leq 10^{-4}$ for others.

\begin{figure*}[t]
    \centering
    \includegraphics[width=0.99\textwidth]{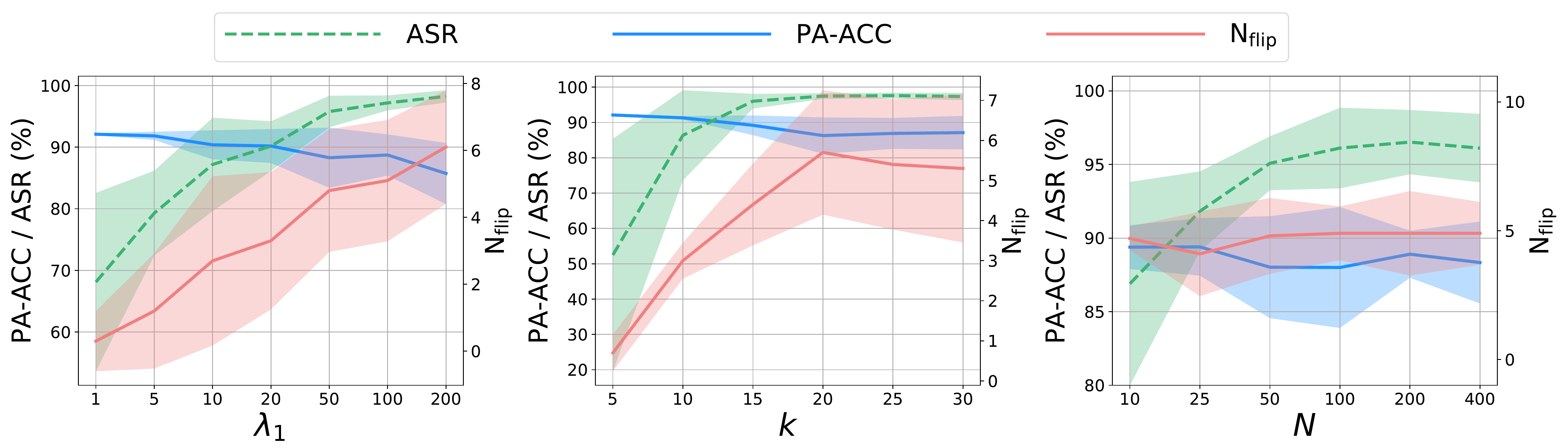}
    \caption{Results of TSA with different parameters $\lambda_1$ ($\lambda_2$ is fixed at 1), $k$, and the number of auxiliary samples $N$ on CIFAR-10. Regions in shadow indicate the standard deviation of attacking into 10 target classes.}
	\label{fig:ablation_study_tsa}
\end{figure*}

\subsubsection{Main Results}
\textbf{Results on CIFAR-10.} The results of TSA and the compared methods are shown in Table \ref{tab:main_tsa}. It shows that FT achieves a higher ASR for VGG, but the accuracy on original samples of the attacked model drop drastically and the $\mathrm{N_{flip}}$ is unacceptable. For the proposed TBT in \cite{rakin2020tbt}, the performance in attacking ResNet is comparable with other methods, but its ASR is too poor for VGG. It can be observed that TSA achieves the least $\mathrm{N_{flip}}$ (less than 7) for all the bit-widths and architectures, which can be attributed to the proposed constraint on the number of flipped bits (see Eq. \ref{eq:tsa_obj}). Besides, TSA can also achieve the highest PA-ACC and a satisfactory ASR (greater than 94\%) in all cases. All results verify the superiority of TSA compared with FT and TBT.

\textbf{Results on ImageNet.}
Table \ref{tab:defense_tsa} shows the results on ImageNet. Because there is no limitation on the number of flipped bits, the FT achieves most $\mathrm{N_{flip}}$ among three methods in all cases. TBT shows its very competitive performance in attacking both architectures. However, our method can balance three metrics better, $i.e.$, TSA can achieve a comparable PA-ACC, the highest ASR, and the least $\mathrm{N_{flip}}$. Especially in attacking 8-bits quantized ResNet, TSA improves greatly in terms of $\mathrm{N_{flip}}$ with the highest PA-ACC and ASR. For TSA, we also observe that the $\mathrm{N_{flip}}$ in attacking VGG is significantly more than that in attacking ResNet. This finding can further verify the effect of the model width against the bit-flip based attack.

\begin{figure}[t]
    \centering
    \subfigure{
    \includegraphics[width=0.15\textwidth]{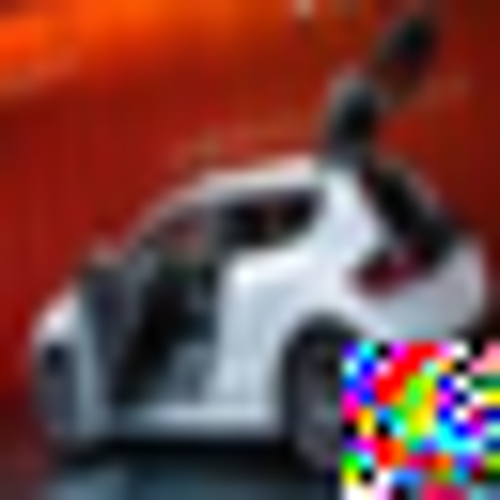}}
    \hspace{5mm}
    \subfigure{
    \includegraphics[width=0.15\textwidth]{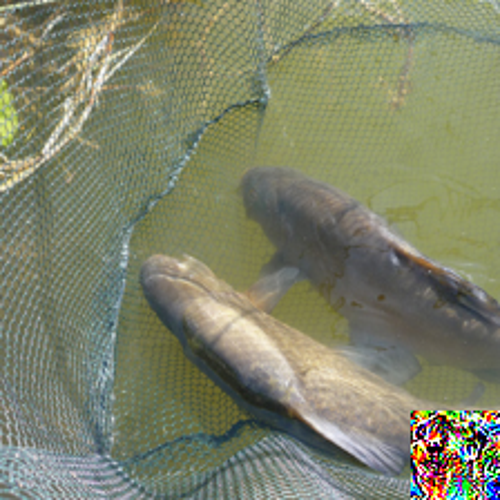}}
    \caption{Visualization of samples embedded with a trigger generated by TSA. Left: an image from CIFAR-10 with 10$\times$10 trigger. Right: an image from ImageNet with 40$\times$40 trigger. }
	\label{fig:trigger_vis_tsa}
\end{figure}

\subsubsection{Resistance to Defense Method}
\textbf{Resistance to Piece-wise Clustering.}
We evaluate three methods in attacking models training with the piece-wise clustering and keep all settings as those in Section \ref{sec:set_tsa}. The results are shown in Table \ref{tab:defense_tsa}. It shows that the defense is effective against the bit-flip based attack, resulting in a lower ASR or a more $\mathrm{N_{flip}}$. On CIFAR-10, the ASR of FT and TBT is less than 50\%. Besides, it is worth noting that the PA-ACC of the attacked model on ImageNet for the FT and TBT is only 0.10 \%, $i.e.$, the target model is converted into a random output generator. It also indicates the effectiveness of the piece-wise clustering defense. However, our TSA can still achieve both a satisfactory PA-ACC and ASR with the least $\mathrm{N_{flip}}$ in all cases. These results demonstrate that the proposed method can still identify critical bits and generate a trigger effectively, even though model training with the piece-wise clustering is more robust.

\textbf{Resistance to Larger Model Capacity.}
The results of attacking the models with 2$\times$ width are presented in Table \ref{tab:defense_tsa}. The increased $\mathrm{N_{flip}}$ in all cases illustrates that larger model capacity can improve the model robustness against these three attacks to some extent. Among three methods, TSA can achieve the least $\mathrm{N_{flip}}$ and $\Delta \mathrm{N_{flip}}$ with the highest PA-ACC and ASR on both datasets. Comparing the results of two defense methods on ImageNet, we find that the piece-wise clustering can provide more effective defense than the larger model capacity, which is consistent with the finding in Section \ref{sec:defense_ssa}.

\subsubsection{Ablation Studies}
We investigate the effect of the $\lambda_1$ ($\lambda_2$ is fixed at 1), $k$, and the number of auxiliary samples $N$ for the proposed TSA. We show the results on CIAFR-10 using 8-bit quantized ResNet in Fig. \ref{fig:ablation_study_tsa} under various values of $\lambda_1$ while $k$ is 20, and various values of $k$ while $\lambda_1$ is 100. We also study different size of the auxiliary sample set from 10 to 400 and keep other settings as those in Section \ref{sec:set_tsa}.
It can be seen from the figure that larger $\lambda_1$ encourages a higher ASR but a lower PA-ACC and more bit-flips, which is in agreement with the designed objective function in Eq. \ref{eq:tsa_obj}. When $\lambda_1$ is selected from 20 to 100, TSA can achieve a relatively better trade-off between the three metrics. The plot of parameter $k$ presents that $\mathrm{N_{flip}}$ increases with the increase of $k$ when $k$ is less than 20, which illustrates the effectiveness of the proposed constraint on the number of flipped bits. We also find that $k=15$ is a feasible setting in attacking 8-bit quantized ResNet on CIAFR-10. As shown in the plot of parameter $N$, a larger number of auxiliary samples leads to a higher ASR when $N$ is less than 200, while has little effect on PA-ACC and $\mathrm{N_{flip}}$. The rationale behind this finding may be that more auxiliary samples make a better generalization of the generated trigger and flipped bits on the unseen samples. 

\begin{table}[]
\caption{Results of TSA with various trigger sizes against 8-bit quantized ResNet on CIFAR-10. The mean and standard deviation of all metrics calculated by attacking into 10 target classes.}
\centering
\begin{tabular}{cccc}
\hline
\begin{tabular}[c]{@{}c@{}}Trigger Size\end{tabular} & \begin{tabular}[c]{@{}c@{}}PA-ACC (\%)\end{tabular} & \begin{tabular}[c]{@{}c@{}}ASR (\%)\end{tabular} & $\mathrm{N_{flip}}$ \\ \hline
6$\times$6 & 51.84\scriptsize{$\pm$17.87} & 95.02\scriptsize{$\pm$3.94} & 14.9\scriptsize{$\pm$3.30} \\
8$\times$8 & 75.40\scriptsize{$\pm$6.90} & 95.18\scriptsize{$\pm$3.03} & 9.6\scriptsize{$\pm$1.56} \\
10$\times$10 & 88.82\scriptsize{$\pm$2.73} & 95.67\scriptsize{$\pm$2.32} & 4.3\scriptsize{$\pm$1.49} \\
12$\times$12 & 91.43\scriptsize{$\pm$1.29} & 98.61\scriptsize{$\pm$1.39} & 1.5\scriptsize{$\pm$1.91} \\
14$\times$14 & 92.00\scriptsize{$\pm$0.47} & 99.82\scriptsize{$\pm$0.17} & 0.4\scriptsize{$\pm$1.20} \\
16$\times$16 & 92.15\scriptsize{$\pm$0.02} & 99.99\scriptsize{$\pm$0.01} & 0.1\scriptsize{$\pm$0.30} \\ \hline
\end{tabular}
\label{tab:trigger_size}
\end{table}

\begin{table}[]
\caption{Results of TSA with different trigger locations against 8-bit quantized ResNet on CIFAR-10. The mean and standard deviation of all metrics calculated by attacking into 10 target classes.}
\centering
\begin{tabular}{cccc}
\hline
\begin{tabular}[c]{@{}c@{}}Trigger Location\end{tabular} & \multicolumn{1}{c}{PA-ACC (\%)} & \multicolumn{1}{c}{ASR (\%)} & \multicolumn{1}{c}{$\mathrm{N_{flip}}$} \\ \hline
\textit{bottom left} & 88.93\scriptsize{$\pm$3.43} & 96.62\scriptsize{$\pm$1.83} & 4.0\scriptsize{$\pm$2.49} \\
\textit{bottom right} & 88.92\scriptsize{$\pm$2.38} & 96.46\scriptsize{$\pm$2.16} & 4.7\scriptsize{$\pm$0.90} \\
\textit{top left} & 90.81\scriptsize{$\pm$1.26} & 98.56\scriptsize{$\pm$1.13} & 3.2\scriptsize{$\pm$2.18} \\
\textit{top right} & 90.86\scriptsize{$\pm$1.45} & 96.50\scriptsize{$\pm$2.47} & 3.0\scriptsize{$\pm$1.55} \\ \hline
\end{tabular}
\label{tab:trigger_loc}
\end{table}

\subsubsection{Trigger Design}
We delve into the trigger design for the TSA in this part. We firstly visualize the generated trigger with the above settings in Fig. \ref{fig:trigger_vis_tsa}. The visualization shows that the trigger area is relatively small in the whole image, especially for the sample from ImageNet, resulting in a stealthy attack.
We also investigate how the trigger size and location influence the attack effectiveness in this part. We conduct experiments using various trigger sizes and keep other settings in line with Section \ref{sec:set_tsa}. The results are shown in Table \ref{tab:trigger_size}. With the increase of the trigger size, the attack performance is improved overall, including a higher PA-ACC, a higher ASR, and a less $\mathrm{N_{flip}}$. Meanwhile, a larger trigger size means more modification on the sample, leading to poor stealthiness. The results of TSA with different trigger locations are shown in Table \ref{tab:trigger_loc}. It shows that TSA can achieve satisfactory attack performance in all cases. However, an interesting observation is that TSA achieves a relatively better result when the trigger is located at the top left. This may be because the important features on CIFAR-10 are mainly near the top left, which is consistent with the finding in \cite{rakin2020tbt}.

\subsection{A Closer Look at the Proposed Method}

\subsubsection{Effect of the Target Class}

We study the effect of the target class for the proposed method in this section. Table \ref{tab:ssa_class} shows the results of SSA varying target class on CIAFR-10, where we attack 100 samples from other classes into each target class. For different target class, SSA always achieves a 100.0\% ASR and about 88\% PA-ACC with a few bit-flips. It verifies that different target class has a little effect on the attack performance for SSA. The results of TSA are shown in Table \ref{tab:tsa_class}. For the TSA, an interesting finding is that the attack performance changes a lot across target classes, which is different from SSA. Therefore, the difficulty of attacking different target classes is different. For example, TSA achieves a relatively high PA-ACC and ASR with only 2 bit-flips in attacking into class 2; the PA-ACC is only 78.30\% with the highest bit-flips in attacking into class 5.

\begin{table}[t]
\caption{Results of SSA with different target classes in attacking 8-bit quantized ResNet on CIFAR-10. The mean and standard deviation of PA-ACC and $\mathrm{N_{flip}}$ are calculated by attacking the 100 images.}
\centering
\begin{tabular}{cccc}
\hline
Target Class & PA-ACC (\%) & ASR (\%) & $\mathrm{N_{flip}}$ \\ \hline
0 & 87.95\scriptsize{$\pm$2.66} & 100.0 & 5.61\scriptsize{$\pm$1.62} \\
1 & 88.49\scriptsize{$\pm$2.48} & 100.0 & 5.61\scriptsize{$\pm$1.62} \\
2 & 88.22\scriptsize{$\pm$2.81} & 100.0 & 5.61\scriptsize{$\pm$1.62} \\
3 & 87.99\scriptsize{$\pm$2.76} & 100.0 & 5.61\scriptsize{$\pm$1.62} \\
4 & 87.80\scriptsize{$\pm$2.73} & 100.0 & 5.61\scriptsize{$\pm$1.62} \\
5 & 88.18\scriptsize{$\pm$2.61} & 100.0 & 5.61\scriptsize{$\pm$1.62} \\
6 & 88.13\scriptsize{$\pm$2.58} & 100.0 & 5.56\scriptsize{$\pm$1.56} \\
7 & 88.26\scriptsize{$\pm$2.69} & 100.0 & 5.56\scriptsize{$\pm$1.56} \\
8 & 88.55\scriptsize{$\pm$2.43} & 100.0 & 5.51\scriptsize{$\pm$1.50} \\
9 & 88.45\scriptsize{$\pm$2.54} & 100.0 & 5.50\scriptsize{$\pm$1.50} \\ \hline
\end{tabular}
\label{tab:ssa_class}
\end{table}

\begin{table}[t]
\caption{Results of TSA with different target classes in attacking 8-bit quantized ResNet on CIFAR-10.}
\centering
\begin{tabular}{cccc}
\hline
Target Class & PA-ACC (\%) & ASR (\%) & $\mathrm{N_{flip}}$ \\ \hline
0 & 89.58 & 96.50 & 5 \\
1 & 91.87 & 96.11 & 2 \\
2 & 92.01 & 97.42 & 2 \\
3 & 80.41 & 96.94 & 5 \\
4 & 90.46 & 94.04 & 4 \\
5 & 78.30 & 98.87 & 7 \\
6 & 91.10 & 92.75 & 4 \\
7 & 89.20 & 97.79 & 5 \\
8 & 90.75 & 91.95 & 5 \\
9 & 87.17 & 98.26 & 6 \\ \hline
\end{tabular}
\label{tab:tsa_class}
\end{table}

\subsubsection{Complexity Analysis}
\label{sec:complexity}
The computational complexity of SSA consists of two parts, the forward and backward pass. 
In terms of the forward pass, since $\bm{\Theta}, \mathbf{B}_{\{1,\ldots, K\}\backslash \{s, t\}}$ are fixed during the optimization, their involved terms, including $g(\bm{x}; \bm{\Theta})$ and $p(\bm{x}; \bm{\Theta}, \mathbf{B}_i)|_{i\neq s,t}$, are calculated only one time. 
The main cost from $\hat{\mathbf{B}}_s$ and $\hat{\mathbf{B}}_t$ is $O(2(N_2+1)C^2Q)$ per iteration, as there are $N_2+1$ samples. 
In terms of the backward pass, the main cost is from the update of $\hat{\bm{b}}^{r+1}$, which is $O(2(N_2+1)CQ)$ per iteration in the gradient descent. 
Since all other updates are very simple, their costs are omitted here. 
Thus, the overall computational cost is $O\big(T_{outer} \cdot T_{inner} \cdot [ 2(N_2+1)CQ \cdot (C+1) ] \big)$, with $T_{outer}$ being the iteration of the overall algorithm and $T_{inner}$ indicating the number of gradient steps in updating $\hat{\bm{b}}^{r+1}$. 
As shown in Section \ref{sec:convergence}, SSA always converges very fast in our experiments, thus $T_{outer}$ is not very large. 
As demonstrated in Section \ref{sec:set_ssa}, $T_{inner}$ is set to 5 in our experiments. 
In short, the proposed method can be optimized very efficiently.

For TSA, since the optimization involves updating the input variable ($i.e.$, the trigger $\bm{q}$), the main costs are the forward process and computing the gradients of $L$ \textit{w.r.t.} $\hat{\bm{b}}$ and $\bm{q}$. Similar to the analysis for SSA, the cost of updating $\hat{\bm{b}}$ per iteration is $O(2KNCQ)$ in the gradient descent. The costs of the forward process and computing the gradients of $L$ \textit{w.r.t.} $\bm{q}$ depends on the attacked model $f$, which are the same with the complexities of the forward and backward pass in the standard model training. Moreover, considering the very small auxiliary sample set (128 on CIFAR-10 and 512 on ImageNet), TSA could be optimized with an acceptable time.

\begin{figure}[t]
    \centering
    \subfigure[SSA on CIFAR-10]{
    \includegraphics[width=0.22\textwidth,height=30mm]{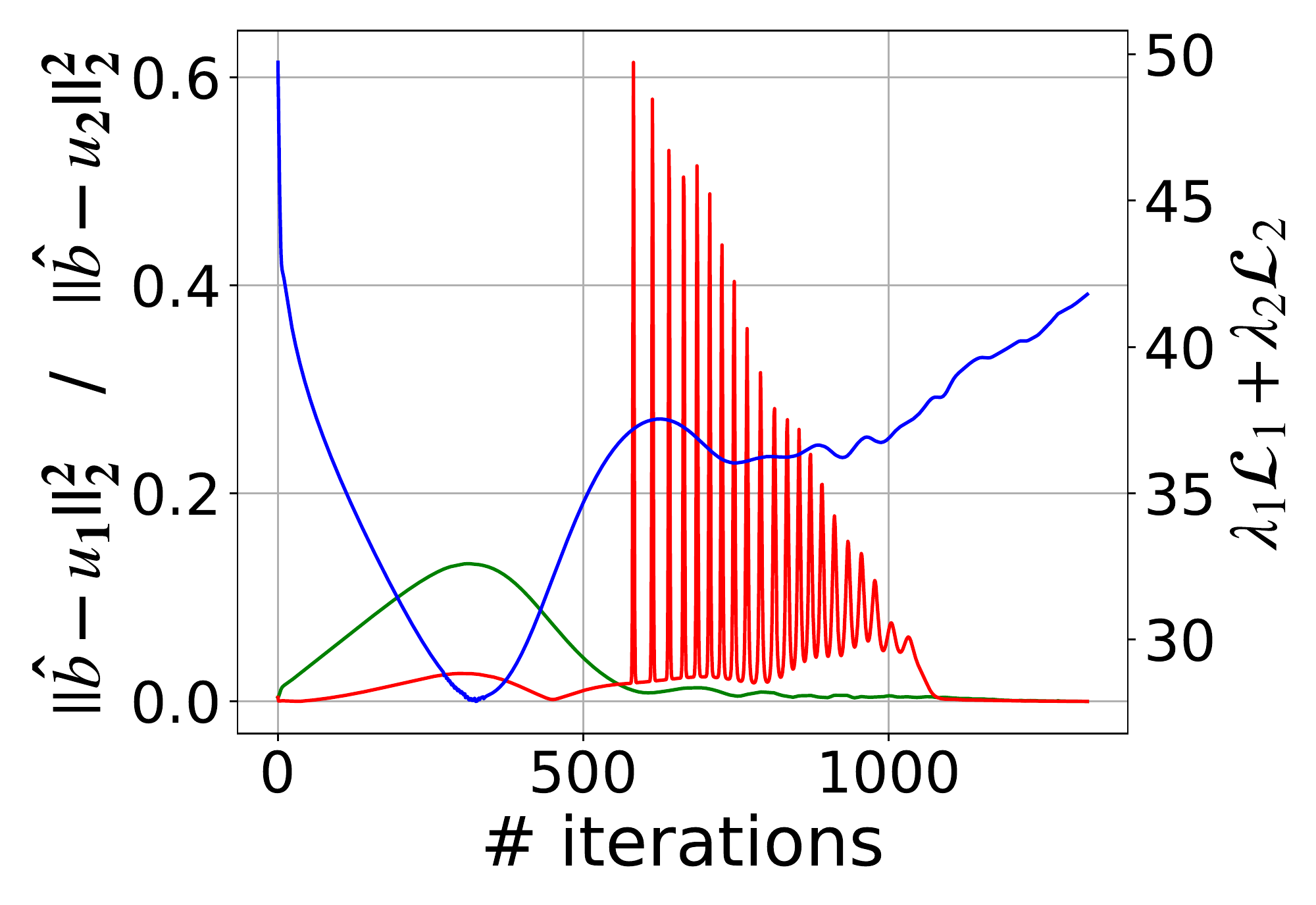}}
    \subfigure[SSA on ImageNet]{
    \includegraphics[width=0.25\textwidth,height=30mm]{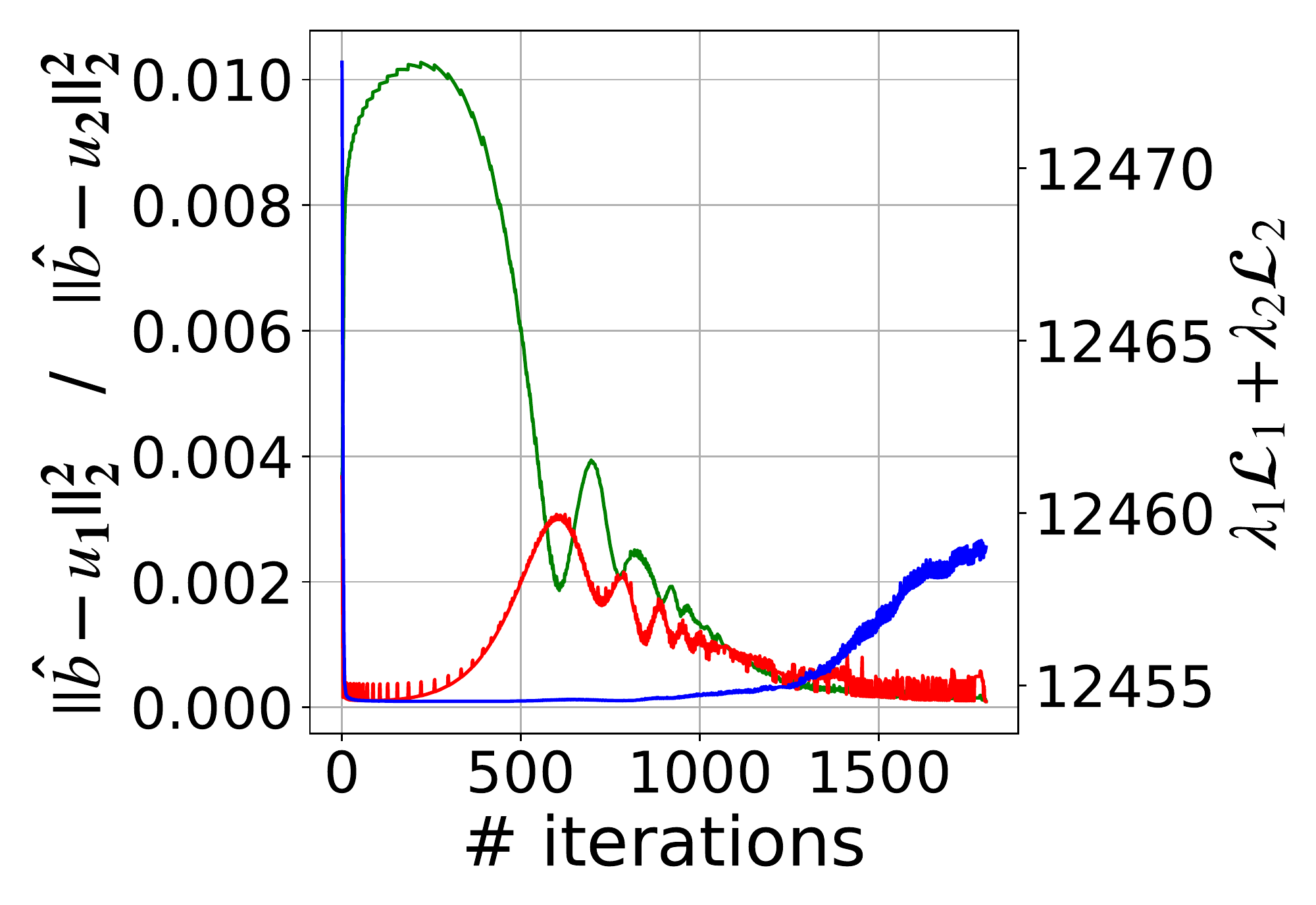}}
    \subfigure[TSA on CIFAR-10]{
    \includegraphics[width=0.23\textwidth,height=30mm]{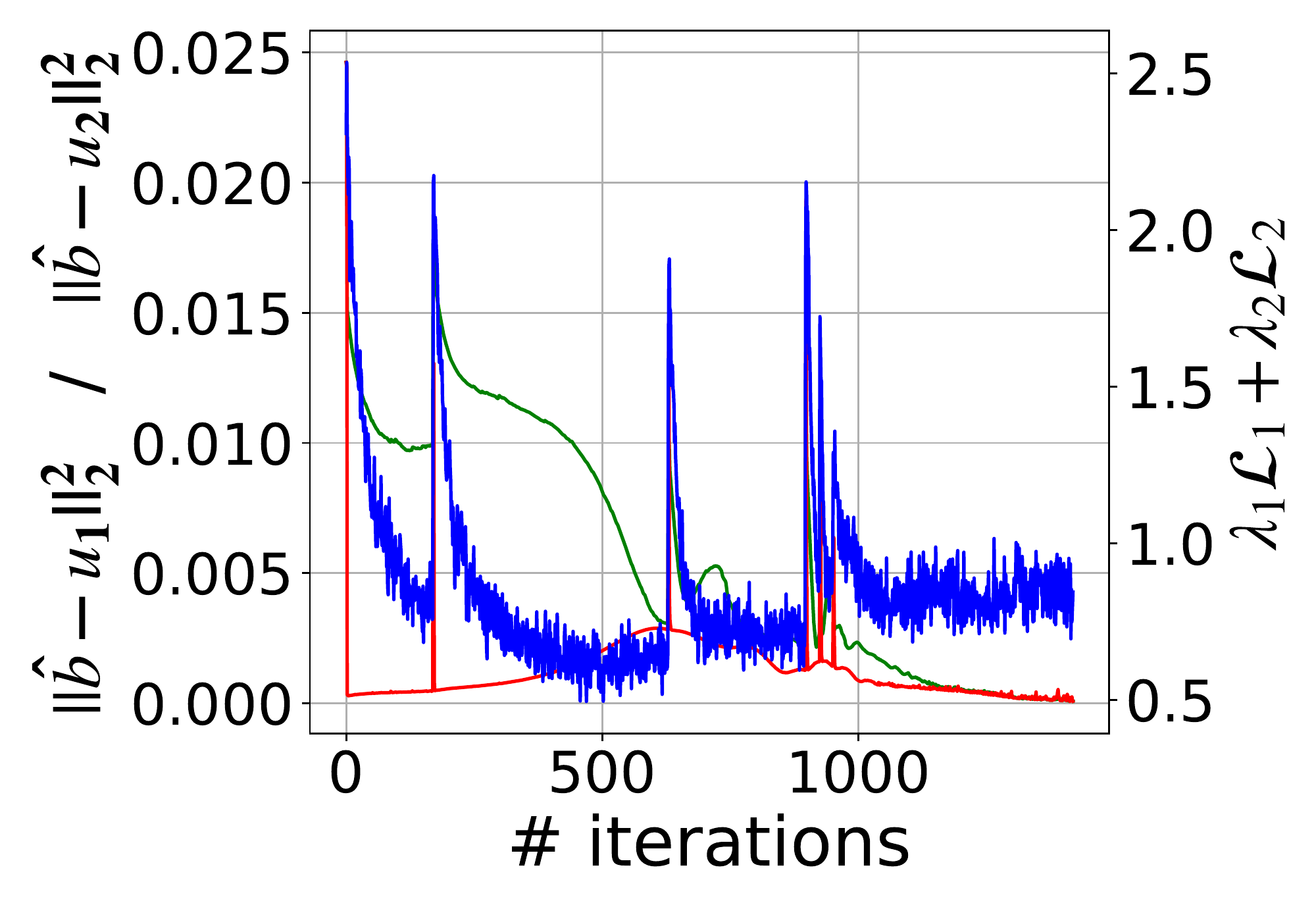}}
    \subfigure[TSA on ImageNet]{
    \includegraphics[width=0.23\textwidth,height=30mm]{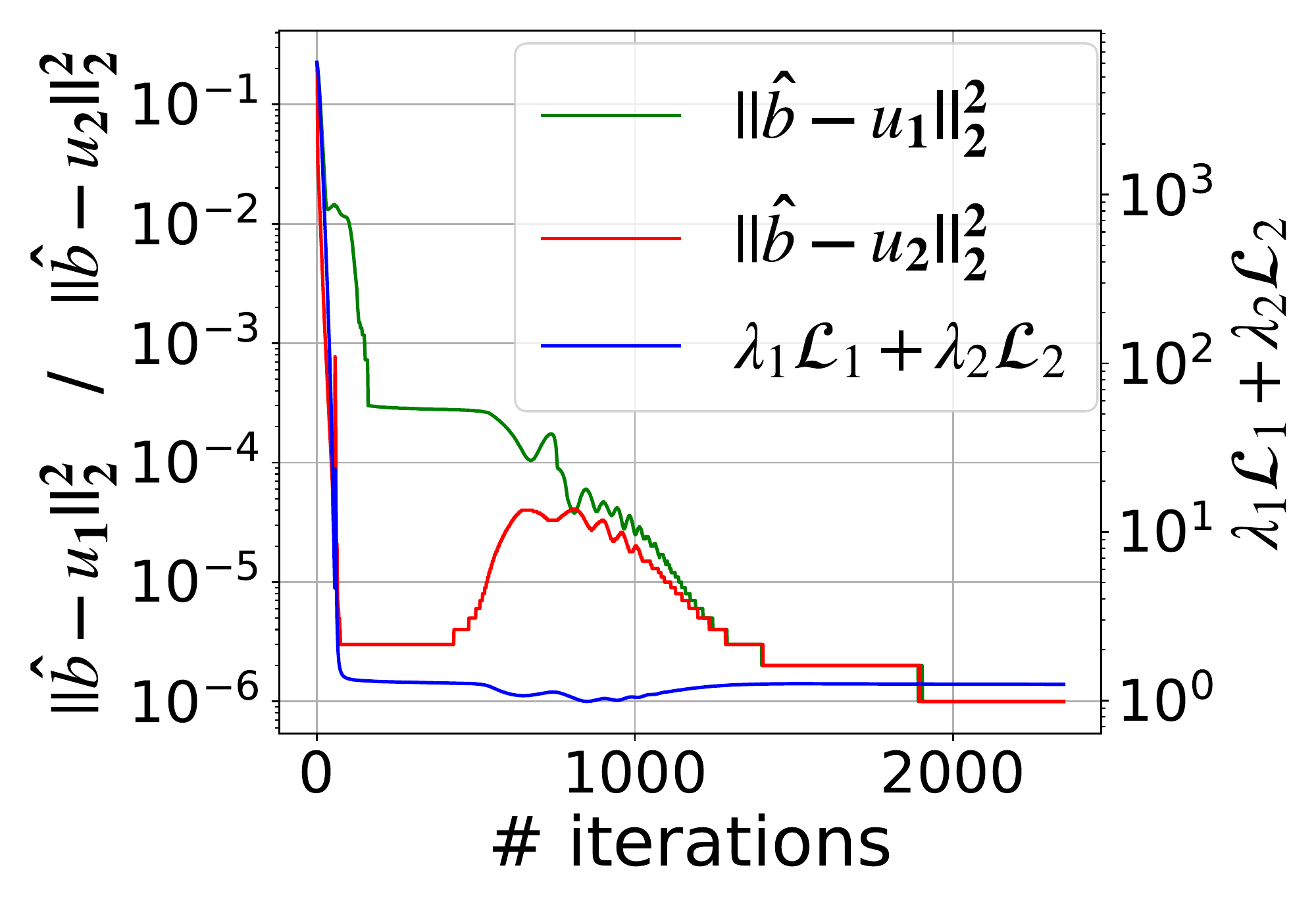}}
    \caption{Numerical convergence analysis of SSA and TSA on CIFAR-10 and ImageNet, respectively. We present the values of $|| \hat{\bm{b}}-\bm{u}_1 ||^2_2 $, $|| \hat{\bm{b}}-\bm{u}_2 ||^2_2 $ and $\lambda_2 \mathcal{L}_1+\lambda_2 \mathcal{L}_2$ at different iterations in attacking 8-bit quantized ResNet. Note that the maximum number of iterations is set to 2,000 and 3,000 for SSA and TSA, respectively.}
	\label{fig:convergence}
\end{figure}

\subsubsection{Numerical  Convergence Analysis}
\label{sec:convergence}
We present the numerical convergence of SSA and TSA in Fig. \ref{fig:convergence}. 
Note that $||  \hat{\bm{b}}- \bm{u}_1||_2^2$ and $||  \hat{\bm{b}}- \bm{u}_2||_2^2$ characterize the degree of satisfaction of the box and $\ell_2$-sphere constraint, respectively. $\lambda_1 \mathcal{L}_1+\lambda_2 \mathcal{L}_2$ means the attack performance to some extent, where a lower value corresponds to a better attack.

For the two examples of SSA on CIFAR-10 and ImageNet, the values of both indicators first increase, then drop, and finally close to 0. 
Another interesting observation is that $\lambda_1 \mathcal{L}_1+\lambda_2 \mathcal{L}_2$ first decreases evidently and then increases slightly. 
Such findings illustrate the optimization process of SSA. 
In the early iterations, modifying the model parameters tends to achieve the effectiveness and stealthiness goals; in the late iterations, $\hat{\bm{b}}$ is encouraged to satisfy the box and $l_2$-sphere constraint. 
We also observe that both examples stop when meeting $||  \hat{\bm{b}}- \bm{u}_1||_2^2 \leq 10^{-4}$ and $||  \hat{\bm{b}}- \bm{u}_2||_2^2 \leq 10^{-4}$ and do not exceed the maximum number of iterations.

For the TSA, $||  \hat{\bm{b}}- \bm{u}_1||_2^2$ and $||  \hat{\bm{b}}- \bm{u}_2||_2^2$ start from relatively high values and could be optimized to be lower than the stop threshold. At the end of the optimization, TSA can modify the model parameters and generate a trigger to achieve a low value of $\lambda_1 \mathcal{L}_1+\lambda_2 \mathcal{L}_2$ on both datasets. Similar to the SSA, the optimization can stop within the maximum number of iterations.
All the above numerical results demonstrate the fast convergence of our proposed method in practice.

\section{Conclusion and Future Work}
In this work, we have explored a novel attack paradigm that the weights of a deployed DNN can be slightly changed via bit flipping in the memory to achieve some malicious purposes. Firstly, we present the general formulation considering the attack effectiveness and stealthiness and formulate it as a mixed integer programming (MIP) problem since the weights are stored as binary bits in the memory. Based on the general formulation, we propose two types of attack: SSA and TSA.
while the predictions on other samples are not significantly influenced. Furthermore, we solve the MIP problem with an effective and efficient continuous algorithm. 
Since the critical bits are determined through optimization, SSA and TSA can achieve the attack goals by flipping a few bits under different experimental settings. 

Future studies of the bit-flip based attack include, but are
not limited to: \textbf{1)} Specifying the proposed framework as other types of attack with different malicious purposes; \textbf{2)} Extending it to the different tasks, $e.g.$, face recognition and autonomous driving; \textbf{3)} Exploring more strict settings than the whit-box one; \textbf{4)} Using our method to evaluate the parameter robustness of models with different architectures, training strategies, $etc.$, due to its least number of bit-flips. We hope that this work can draw more attention to the security of the deployed DNNs and encourage further investigation on the defense against the bit-flip based attack.

\ifCLASSOPTIONcaptionsoff
  \newpage
\fi



\normalem
\bibliographystyle{IEEEtran}
\bibliography{IEEEabrv,ref}

\appendices

\section{Update $\hat{\bm{b}}$ by Gradient Descent}
Here, we derive the gradient of $L$ \textit{w.r.t.} $\hat{\bm{b}}$, which is adopted to update $\hat{\bm{b}}^{r+1}$ by gradient descent. In the following two parts, we present the derivations for the SSA and TSA, respectively.

\subsection{Update $\hat{\bm{b}}$ for SSA}
For SSA, $\bm{b}$ corresponds to the reshaped and concatenated $\mathbf{B}_s$ and $\mathbf{B}_t$ and $\hat{\bm{b}}$ could be the reshaped and concatenated $\hat{\mathbf{B}}_s$ and $\hat{\mathbf{B}}_t$, respectively. 
Based on the specified objective function for SSA, the derivation consists of the following parts.

\textbf{Derivation of $\partial \mathcal{L}_1^{SSA}(\bm{x}, s, t; \hat{\mathbf{B}}_s, \hat{\mathbf{B}}_t) / \partial \hat{\bm{b}}$.} ~
For clarity, here we firstly repeat some definitions, 

\begin{equation}
\begin{split}
\mathcal{L}_1^{SSA}(\bm{x}, s, t; \hat{\mathbf{B}}_s, \hat{\mathbf{B}}_t) & = \max\big(\tau-p(\bm{x}; \bm{\Theta}, \hat{\mathbf{B}}_t)+\delta, 0\big)   \\ & + \max\big(p(\bm{x}; \bm{\Theta}, \hat{\mathbf{B}}_s)-\tau + \delta, 0\big),
\end{split}
\label{loss:1 appendix}
\end{equation}

\begin{equation}
\begin{split}
p(\bm{x}; \bm{\Theta}, \hat{\mathbf{B}}_i)=[h(\hat{\mathbf{B}}_{i,1});& h(\hat{\mathbf{B}}_{i,2});...;h(\hat{\mathbf{B}}_{i,C})]^{\top}g(\bm{x};\bm{\Theta}),
\end{split}
\label{eq: p appendix}
\end{equation}

\begin{equation}
h(\bm{v})=(-2^{Q-1} \cdot v_Q + \sum_{i=1}^{Q-1}{2^{i-1} \cdot v_i}) \cdot \Delta^l.
\label{eq: h appendix}
\end{equation}

Then, we obtain that 
\begin{equation}
\begin{split}
\frac{\partial p(\bm{x};\bm{\Theta},\hat{\mathbf{B}}_i)}{\partial \hat{\mathbf{B}}_i} = [ & g_1(\bm{x};\bm{\Theta}) \cdot (\frac{\nabla h(\hat{\mathbf{B}}_{i,1})}{\partial \hat{\mathbf{B}}_{i,1}})^\top; ...; \\ & g_C(\bm{x};\bm{\Theta}) \cdot (\frac{\nabla h(\hat{\mathbf{B}}_{i,C})}{\nabla \hat{\mathbf{B}}_{i,C}})^\top],
\end{split}
\label{eq:pb appendix}
\end{equation}
where $\frac{\nabla h(\bm{v})}{\nabla \bm{v}} = [2^0; 2^1, \ldots, 2^{Q-2}; -2^{Q-1}] \cdot \Delta^l$ is a constant, and here $l$ indicates the last layer; $g_j(\bm{x};\bm{\Theta})$ denotes the $j$-th entry of the vector $g(\bm{x};\bm{\Theta})$. 

Utilizing (\ref{eq:pb appendix}), we have 
\begin{equation}
\begin{split}
        &\frac{\partial \mathcal{L}_1^{SSA}(\bm{x},\! s,\! t; \!\hat{\mathbf{B}}_s, \!\hat{\mathbf{B}}_t)}{\partial \hat{\mathbf{B}}_s} \! = \!
        \left\{\begin{array}{l}
        \frac{\partial p(\bm{x};\bm{\Theta},\hat{\mathbf{B}}_s)}{\partial \hat{\mathbf{B}}_s}, \ \textrm{if} \ p(\bm{x}; \bm{\Theta}, \mathbf{B}_s) \!>\!\tau \! - \! \delta \\
        \bm{0}, \ \textrm{otherwise}
        \end{array}\right. \!\!\!,
        \\
        &\frac{\partial \mathcal{L}_1^{SSA}(\bm{x}, \! s, \! t;\! \hat{\mathbf{B}}_s, \!\hat{\mathbf{B}}_t)}{\partial \hat{\mathbf{B}}_t} \! = \!
        \left\{\begin{array}{l}
        -\frac{\partial p(\bm{x};\bm{\Theta},\hat{\mathbf{B}}_t)}{\partial \hat{\mathbf{B}}_t}, \ \textrm{if} \ p(\bm{x}; \bm{\Theta}, \mathbf{B}_t)\!<\!\tau \! + \! \delta \\
        \bm{0}, \ \textrm{otherwise}
        \end{array}
        \right. \!\!\!\! \! .
\end{split}
\end{equation}
Thus, we obtain that 
\begin{equation}
\begin{split}
\frac{\partial \mathcal{L}_1^{SSA}(\bm{x}, \!s, \!t; \!\hat{\mathbf{B}}_s,\! \hat{\mathbf{B}}_t)}{\partial \hat{\bm{b}}} \!= \!
\big[ & \text{Reshape}\big(\frac{\partial\mathcal{L}_1^{SSA}(\bm{x}, s, t; \hat{\mathbf{B}}_s, \hat{\mathbf{B}}_t)}{\partial \hat{\mathbf{B}}_s}\big); \\
& \text{Reshape}\big(\frac{\partial \mathcal{L}_1^{SSA}(\bm{x}, s, t; \hat{\mathbf{B}}_s, \hat{\mathbf{B}}_t)}{\partial \hat{\mathbf{B}}_t}\big) \big],
\end{split}
\label{eq: gradient of L1 to b ssa}
\end{equation}
where $\text{Reshape}(\cdot)$ elongates a matrix to a vector along the column.

\textbf{Derivation of $\partial \mathcal{L}_2(\bm{D}_2;\hat{\mathbf{B}}_s, \hat{\mathbf{B}}_t) / \partial \hat{\bm{b}}$.} ~
For clarity, here we firstly repeat the following definition 

\begin{equation}
\begin{split}
         \mathcal{L}_2(\bm{D}_2;\hat{\mathbf{B}}_s, \hat{\mathbf{B}}_t)=\sum_{(\bm{x}_i,y_i)\in\bm{D}_2}{\ell(f(\bm{x}_i; \hat{\mathbf{B}}_s, \hat{\mathbf{B}}_t), y_i)},
\end{split}
\end{equation}
where $f_j(\bm{x}_i; \hat{\mathbf{B}}_s, \hat{\mathbf{B}}_t)$
indicates the posterior probability of $\bm{x}_i$ $w.r.t.$ class $j$.
$\bm{D}_2=\{(\bm{x}_i, y_i)\}_{i=1}^{N_2}$ denotes the auxiliary sample set.
$\ell(\cdot, \cdot)$ is specified as the cross entropy loss.
Then, we have 
\begin{equation}
\begin{split}
    &\frac{\partial \mathcal{L}_2(\bm{D}_2;\hat{\mathbf{B}}_s, \hat{\mathbf{B}}_t)}{\partial \hat{\mathbf{B}}_s} \\ 
    = &
    \sum_{(\bm{x}_i,y_i)\in\bm{D}_2}\bigg[ 
    \big( \mathbb{I}(y_i=s)  - 
    f_s(\bm{x}_i; \hat{\mathbf{B}}_s, \hat{\mathbf{B}}_t) \big) \cdot \frac{\partial p(\bm{x}_i; \bm{\Theta}, \hat{\mathbf{B}}_s )}{\partial \hat{\mathbf{B}}_s}
    \bigg], 
\end{split}
\end{equation}

\begin{equation}
\begin{split}
    &\frac{\partial \mathcal{L}_2(\bm{D}_2;\hat{\mathbf{B}}_s, \hat{\mathbf{B}}_t)}{\partial \hat{\mathbf{B}}_t} \\
    = &
    \sum_{(\bm{x}_i,y_i)\in\bm{D}_2}\bigg[ 
    \big( \mathbb{I}(y_i=t)  - 
    f_t(\bm{x}_i; \hat{\mathbf{B}}_s, \hat{\mathbf{B}}_t) \big) \cdot \frac{\partial p(\bm{x}_i; \bm{\Theta}, \hat{\mathbf{B}}_t )}{\partial \hat{\mathbf{B}}_t}
    \bigg],
\end{split}
\end{equation}
where $\mathbb{I}(a) = 1$ of $a$ is true, otherwise $\mathbb{I}(a) = 0$. 
Thus, we obtain 
\begin{equation}
\begin{split}
\frac{\partial \mathcal{L}_2(\bm{D}_2;\hat{\mathbf{B}}_s, \hat{\mathbf{B}}_t)}{\partial \hat{\bm{b}}} = 
\bigg[ &\text{Reshape}\big(\frac{\partial \mathcal{L}_2(\bm{D}_2;\hat{\mathbf{B}}_s, \hat{\mathbf{B}}_t)}{\partial \hat{\mathbf{B}}_s}\big); \\ 
&\text{Reshape}\big(\frac{\partial \mathcal{L}_2(\bm{D}_2;\hat{\mathbf{B}}_s, \hat{\mathbf{B}}_t)}{\partial \hat{\mathbf{B}}_t}\big) \bigg].
\end{split}
\label{eq: gradient of L2 to b ssa}
\end{equation}

\textbf{Derivation of $\partial L(\hat{\bm{b}}) / \partial \hat{\bm{b}}$.} ~
According to the augmented Lagrangian function, and utilizing Eqs. (\ref{eq: gradient of L1 to b ssa}) and (\ref{eq: gradient of L2 to b ssa}), we obtain the gradient of $L$ \textit{w.r.t.} $\hat{\bm{b}}$ for SSA, as follows.
\begin{equation}
\begin{split}
     \frac{\partial L(\hat{\bm{b}})}{\partial \hat{\bm{b}}}  = &
    \frac{\partial \mathcal{L}_1^{SSA}(\bm{x}, s, t; \hat{\mathbf{B}}_s, \hat{\mathbf{B}}_t)}{\partial \hat{\bm{b}}} +
    \frac{\partial \mathcal{L}_2(\bm{D}_2;\hat{\mathbf{B}}_s, \hat{\mathbf{B}}_t)}{\partial \hat{\bm{b}}}  
     \\ + & \bm{z}_1 + \bm{z}_2  + \rho_1(\hat{\bm{b}}-\bm{u}_1) + \rho_2(\hat{\bm{b}}-\bm{u}_2) \\ + &
    2 (\hat{\bm{b}}-\bm{b}) \cdot \big[\bm{z}_3 + \rho_3 ||\hat{\bm{b}}-\bm{b}||^2_2-k+u_3 \big].
\end{split}
\end{equation}

\subsection{Update $\hat{\bm{b}}$ for TSA}
For TSA, $\bm{b}$ corresponds to the reshaped $\mathbf{B}$ and $\hat{\bm{b}}$ could be the reshaped $\hat{\mathbf{B}}$. We firstly give the derivation of  $\partial \mathcal{L}_1^{TSA}(\phi(\bm{D};\bm{q}), t;\hat{\mathbf{B}}) / \partial \hat{\bm{b}}$ and $\partial \mathcal{L}_2(\bm{D};\hat{\mathbf{B}}) / \partial \hat{\bm{b}}$, and then the derivation of $\partial L(\hat{\bm{b}}) / \partial \hat{\bm{b}}$. 

\textbf{Derivation of $\partial \mathcal{L}_1^{TSA}(\phi(\bm{D};\bm{q}), t;\hat{\mathbf{B}}) / \partial \hat{\bm{b}}$.} ~ For clarity, here we firstly repeat the following definition.
\begin{equation}
\begin{split}
    & \mathcal{L}_{1}^{TSA}(\phi(\bm{D};\bm{q}), t; \hat{\mathbf{B}})\\ = & \sum_{(\bm{x}_i,y_i)\in\bm{D}}{\ell(f((1-\bm{m}) \otimes  \bm{x}_i + \bm{m} \otimes \bm{q}; \hat{\mathbf{B}}), t)},
\end{split}
\end{equation}
where $f((1-\bm{m}) \otimes  \bm{x}_i + \bm{m} \otimes \bm{q}; \hat{\mathbf{B}})$
indicates the posterior probability of $\bm{x}_i$ with a trigger $w.r.t.$ class $j$.
$\bm{D}=\{(\bm{x}_i, y_i)\}_{i=1}^{N}$ denotes the auxiliary sample set.
$\ell(\cdot, \cdot)$ is specified as the cross entropy loss.

Utilizing Eq. (\ref{eq:pb appendix}), we obtain that
\begin{equation}
\begin{split}
    & \frac{\partial \mathcal{L}_{1}^{TSA}(\phi(\bm{D};\bm{q}), t; \hat{\mathbf{B}})}{\partial \hat{\mathbf{B}}_j} \! \\ = & \!
        \left\{\begin{array}{l}
        \sum_{(\bm{x}_i,y_i)\in\bm{D}}\bigg[ 
    \big( 1 - 
   f((1-\bm{m}) \otimes  \bm{x}_i + \bm{m} \otimes \bm{q}; \hat{\mathbf{B}})) \\
   ~~~~~~~~~~~~~~~~~~~\cdot \frac{\partial p((1-\bm{m}) \otimes  \bm{x}_i + \bm{m} \otimes \bm{q}; \bm{\Theta}, \hat{\mathbf{B}}_j )}{\partial \hat{\mathbf{B}}_j} \bigg], \ \textrm{if} \ t=j\\
        \sum_{(\bm{x}_i,y_i)\in\bm{D}}\bigg[ 
    - 
   f((1-\bm{m}) \otimes  \bm{x}_i + \bm{m} \otimes \bm{q}; \hat{\mathbf{B}}) \\
   ~~~~~~~~~~~~~~~~~~~\cdot \frac{\partial p((1-\bm{m}) \otimes  \bm{x}_i + \bm{m} \otimes \bm{q}; \bm{\Theta}, \hat{\mathbf{B}}_j )}{\partial \hat{\mathbf{B}}_j} \bigg], \ \textrm{if} \ t \neq j
        \end{array}\right. \!\!\!.
\end{split}
\end{equation}
Then, we have
\begin{equation}
\begin{split}
    & \frac{\partial \mathcal{L}_{1}^{TSA}(\phi(\bm{D};\bm{q}), t; \hat{\mathbf{B}})}{\partial \hat{\mathbf{B}}} \\ = &
    \bigg[\frac{\partial \mathcal{L}_{1}^{TSA}(\phi(\bm{D};\bm{q}), t; \hat{\mathbf{B}})}{\partial \hat{\mathbf{B}}_0};...;
    \frac{\partial \mathcal{L}_{1}^{TSA}(\phi(\bm{D};\bm{q}), t; \hat{\mathbf{B}})}{\partial \hat{\mathbf{B}}_K}
    \bigg].
\end{split}
\end{equation}
Thus, we have
\begin{equation}
\begin{split}
    \frac{\partial  \mathcal{L}_{1}^{TSA}(\phi(\bm{D};\bm{q}), t; \hat{\mathbf{B}}))}{\partial \hat{\bm{b}}} = \text{Reshape} \bigg[\frac{\partial \mathcal{L}_{1}^{TSA}(\phi(\bm{D};\bm{q}), t; \hat{\mathbf{B}})}{\partial \hat{\mathbf{B}}}\bigg],
\end{split}
\label{eq: gradient of L1 to b tsa}
\end{equation}
where $\textrm{Reshape}(\cdot)$ flattens a 3-D tensor to a vector.

\textbf{Derivation of $\partial \mathcal{L}_2(\bm{D};\hat{\mathbf{B}}) / \partial \hat{\bm{b}}$.} ~~ For clarity, here we firstly repeat the following definition.
\begin{equation}
\begin{split}
    \mathcal{L}_{2}(\bm{D};\hat{\mathbf{B}})=\sum_{(\bm{x}_i,y_i)\in\bm{D}}{\ell(f(\bm{x}_i;\hat{\mathbf{B}}), y_i)}.
\end{split}
\end{equation}

Utilizing Eq. (\ref{eq:pb appendix}), we obtain that
\begin{equation}
\begin{split}
    &\frac{\partial \mathcal{L}_2(\bm{D};\hat{\mathbf{B}})}{\partial \hat{\mathbf{B}}_j} \\ 
    = &
    \sum_{(\bm{x}_i,y_i)\in\bm{D}}\bigg[ 
    \big( \mathbb{I}(y_i=j)  - 
   f(\bm{x}_i; \hat{\mathbf{B}})) \cdot \frac{\partial p(\bm{x}_i; \bm{\Theta}, \hat{\mathbf{B}}_j )}{\partial \hat{\mathbf{B}}_j}
    \bigg].
\end{split}
\end{equation}
Then, we have
\begin{equation}
\begin{split}
    &\frac{\partial \mathcal{L}_2(\bm{D};\hat{\mathbf{B}})}{\partial \hat{\mathbf{B}}} =
    \bigg[ \frac{\partial \mathcal{L}_2(\bm{D};\hat{\mathbf{B}})}{\partial \hat{\mathbf{B}}_j};...;
    \frac{\partial \mathcal{L}_2(\bm{D};\hat{\mathbf{B}})}{\partial \hat{\mathbf{B}}_K}
    \bigg].
\end{split}
\end{equation}
Thus, we have
\begin{equation}
\begin{split}
    \frac{\partial \mathcal{L}_2(\bm{D}_2;\hat{\mathbf{B}})}{\partial \hat{\bm{b}}} = \text{Reshape} \bigg[ 
    \frac{\partial \mathcal{L}_2(\bm{D};\hat{\mathbf{B}})}{\partial \hat{\mathbf{B}}}
    \bigg].
\end{split}
\label{eq: gradient of L2 to b tsa}
\end{equation}

\textbf{Derivation of $\partial L(\hat{\bm{b}}) / \partial \hat{\bm{b}}$.} ~ According to the augmented Lagrangian function, and utilizing Eqs. (\ref{eq: gradient of L1 to b tsa}) and (\ref{eq: gradient of L2 to b tsa}), we obtain the gradient of $L$ \textit{w.r.t.} $\hat{\bm{b}}$ for TSA, as follows.
\begin{equation}
\begin{split}
     \frac{\partial L(\hat{\bm{b}})}{\partial \hat{\bm{b}}}  = &
    \frac{\partial  \mathcal{L}_{1}^{TSA}(\phi(\bm{D};\bm{q}), t; \hat{\mathbf{B}}))}{\partial \hat{\bm{b}}} +
    \frac{\partial \mathcal{L}_2(\bm{D}_2;\hat{\mathbf{B}})}{\partial \hat{\bm{b}}}  
     \\ + & \bm{z}_1 + \bm{z}_2  + \rho_1(\hat{\bm{b}}-\bm{u}_1) + \rho_2(\hat{\bm{b}}-\bm{u}_2) \\ + &
    2 (\hat{\bm{b}}-\bm{b}) \cdot \big[\bm{z}_3 + \rho_3 ||\hat{\bm{b}}-\bm{b}||^2_2-k+u_3 \big].
\end{split}
\end{equation}

\end{document}